\documentclass[aip,cha,
reprint,
secnumarabic,
nofootinbib, tightenlines,
nobibnotes, showkeys, showpacs,
superscriptaddress,
]{revtex4-1}

\newcommand{\ssp}{\ensuremath{\mathcal{M}}}
\newcommand{\sspRed}{\ensuremath{\hat{\ssp}}}
\newcommand{\svec}{\ensuremath{\xi}}
\newcommand{\svecproj}{\ensuremath{\tilde{\xi}}}
\newcommand{\svecRed}{\ensuremath{\hat{\svec}}}
\newcommand{\zeit}{\ensuremath{t}}
\newcommand{\symb}{\ensuremath{\sigma}}
\newcommand{\DataSet}{\ensuremath{\Xi}}
\newcommand{\PerD}{\ensuremath{{\rm PD}}}
\newcommand{\PDelm}{\ensuremath{\mu}}
\newcommand{\po}{\ensuremath{\overline{{\rm po}}}}
\newcommand{\distShadow}{\ensuremath{S}}
\newcommand{\shadowDist}{shadowing distance}

\newcommand{\resol}{\ensuremath{r}}

\newcommand{\LieEl}{\ensuremath{g}}  

\newcommand{\inprod}[2]{\left\langle #1 ,\, #2 \right\rangle}
\newcommand{\fFslice}{first Fourier mode slice}

\newcommand\flowRed[2]{{\hat{f}^{#1}(#2)}}
\newcommand{\Fu}{\tilde{u}}
\newcommand{\SymRed}{\ensuremath{\mathcal{R}}}

\newcommand{\vel}{\ensuremath{v}}

\newcommand{\KS}{Ku\-ra\-mo\-to--Siva\-shin\-sky}
\newcommand{\KSe}{Ku\-ra\-mo\-to--Siva\-shin\-sky equation}

\renewcommand\Im{\ensuremath{{\rm Im}\,}}
\renewcommand\Re{\ensuremath{{\rm Re}\,}}

\newcommand{\beq}{\begin{equation}}
\newcommand{\continue}{\nonumber \\ }

\newcommand{\eeq}{\end{equation}}
\newcommand{\ee}[1] {\label{#1} \end{equation}}

\newcommand{\bea}{\begin{eqnarray}}
\newcommand{\eea}{\end{eqnarray}}
\newcommand{\barr}{\begin{array}}
\newcommand{\earr}{\end{array}}

\renewcommand\Im{\ensuremath{{\rm Im}\,}}
\renewcommand\Re{\ensuremath{{\rm Re}\,}}

\newcommand{\etal}{{\em et al.}}    
\newcommand{\ie}{{i.e.}}        

\newcommand{\rf}     [1] {~\cite{#1}}

\newcommand{\refref} [1] {ref.~\cite{#1}}

\newcommand{\refrefs}[1] {refs.~\cite{#1}}

\newcommand{\refeq}  [1] {(\ref{#1})}

\newcommand{\reffig} [1] {figure~\ref{#1}}

\newcommand{\refFig} [1] {Figure~\ref{#1}}

\newcommand{\refsect}[1] {section~\ref{#1}}

\newcommand{\refappe}[1] {appendix~\ref{#1}}

\newcommand\flow[2]{{f^{#1}(#2)}}
\newcommand\flowKS[2]{{f_{KS}^{#1}(#2)}}
\newcommand\timeflow{{f^t}}
\newcommand{\reals}{\mathbb{R}}
\newcommand{\PoincM}{{\cal P}}      

\renewcommand{\reffig} [1] {Fig.~\ref{#1}}

\renewcommand{\refFig} [1] {Fig.~\ref{#1}}

\renewcommand{\refref} [1] {Ref.~\onlinecite{#1}}

\renewcommand{\refrefs}[1] {Refs.~\onlinecite{#1}}

\newcommand{\subgraphicw}[3]{
    \subfloat[(#2)]{\includegraphics[width= #1 \linewidth]{#3}}
}
\newcommand{\subgraphich}[3]{
        \subfloat[(#2)]{\includegraphics[height= #1 \linewidth]{#3}}
}
\newcommand{\graphicw}[2]{
    \includegraphics[width=#1 \linewidth]{#2}
}

\usepackage{natbib}
\usepackage{amsmath,amsfonts,amssymb,amsbsy,amscd,amsgen}
\usepackage{dcolumn}
\usepackage{bm}
\usepackage[pdftex]{graphicx}
\usepackage[caption=false,justification=raggedright,singlelinecheck=false,labelformat=empty]{subfig}

\usepackage{fancyhdr}
\graphicspath{{figs/}}
\usepackage[pdftex,colorlinks]{hyperref} 

\usepackage[T1]{fontenc}
\usepackage[utf8]{inputenc}
\usepackage{color}
\usepackage[normalem]{ulem} 

\begin{document}
\title[]{Inferring symbolic dynamics of chaotic flows from persistence}

\newcommand{\perShadow}{state space persistence analysis}
\newcommand{\PerShadow}{State space persistence analysis}

\author{G\"okhan Yaln\i z}%
\affiliation{
    Physics Department, 
    Bo\u{g}azi\c{c}i University,
    34342 Istanbul, Turkey
}
\altaffiliation[Current address: ]{IST Austria, 3400 Klosterneuburg, Austria}

\author{Nazmi Burak Budanur}
\affiliation{Nonlinear Dynamics and Turbulence Group,
             IST Austria,
             3400 Klosterneuburg, Austria}
\email{burak.budanur@ist.ac.at}                

\date{\today}

\begin{abstract}
We introduce ``\perShadow{}'' for deducing the 
symbolic dynamics 
of time series data obtained from high-dimensional chaotic 
attractors. 
To this end, we adapt a topological data analysis technique known 
as persistent 
homology for the characterization of state space projections of chaotic 
trajectories and periodic orbits. 
By comparing the shapes along a chaotic trajectory to those of the periodic 
orbits, \perShadow{} quantifies the shape similarity of  
chaotic trajectory segments and the periodic orbits. 
We demonstrate the method by applying it to the three-dimensional R\"{o}ssler 
system and a thirty-dimensional discretization of the \KS{}
partial differential equation in $(1+1)$ dimensions. 
\end{abstract}

\maketitle

\begin{quotation}
One way of studying chaotic attractors systematically is through 
their symbolic 
dynamics, in which one partitions the state space into qualitatively different 
regions and assigns a symbol to each 
such region.\rf{Devaney1992,ASY1997,DasBuch} 
This yields a ``coarse-grained'' state space of 
the system, which can then be reduced to a Markov 
						chain encoding
all possible transitions between the states of the system. 
While it is possible to obtain the symbolic dynamics of 
low-dimensional chaotic 
systems with standard tools such as Poincar\'e maps, when applied to 
high-dimensional systems such as turbulent flows, these tools alone are not 
sufficient to determine symbolic dynamics.\rf{CviGib10,WFSBC15} 
In this paper, we develop ``\perShadow{}'' and demonstrate that
	  it can be utilized to infer the symbolic dynamics 
  	  in very high-dimensional settings.
\end{quotation}

\section{Introduction}
\label{s-intro}

One of the defining features of chaos is the sensitive dependence on 
initial conditions,\rf{Devaney1992,ASY1997,strogb} 
which is a statement of the exponential 
amplification of noise under chaotic dynamics.
The practical corollary of this fundamental property is that any 
prediction based on integrating equations of motion of a chaotic 
system starting 
from an initial condition is exponentially wrong in time since 
all measurements 
come with noise. 
Thus, even with the advanced computing technologies of our day, the 
question of 
``What is the future state of a chaotic system based on a 
measurement of its current state?'' can only be answered for a finite time 
horizon. 
A different and more tractable question is the following: 
What are the possible future states of a chaotic system given an approximate 
measurement of its current state? 

The answer to this question begins with
	a ``coarse-graining'' of the system's state space into 
	regions
with qualitative differences, 
	followed by determining the transition rules between these 
	regions.
The associated methods of the dynamical systems theory are 
known as symbolic dynamics.\rf{Devaney1992,ASY1997,DasBuch} 
While these techniques lie at the heart of some of the 
most fundamental 
results
of chaos theory such as 
Smale's proof\rf{smale1965,smale} of the 
		     Birkhoff--Smale theorem,\rf{Birkhoff1935}
existing symbolic 
dynamics methods can only be applied to low-dimensional systems, namely 
the ones 
that can be
effectively described by one- or two-dimensional maps. 

Some examples of continuous-time chaos, 
such as the Lorenz\rf{lorenz63} 
and R\"ossler\rf{ross} systems,
can be reduced to one-dimensional return maps by means of 
Poincar\'e sections.\rf{DV03,LDM1995}
This is possible because 
both models are three-dimensional with
a single positive Lyapunov exponent, which yields a ``thin'' 
attractor due to strong contraction in 
the direction pointing 
outwards from the attractor.\rf{strogb,DasBuch} 
Many real-life examples of chaos, in contrast, take 
place in systems with many ($D \gg 3$) degrees of freedom. 
Examples include 
fluid turbulence,\rf{focusPOT}
cardiac dynamics,\rf{ChaosCardiacFocus2017}
and evolution.\rf{DI2014} 
Generally, such systems cannot be reduced to
low-dimensional maps, 
except in special cases close to the onset of chaos.\rf{KreEck12} 
However, the observations based on computer 
simulations\rf{LucKer14,CviGib10,WFSBC15} suggest that 
high-dimensional systems 
such as turbulent flows exhibit a large catalog of motions
that can be 
associated with the time-periodic solutions of the governing equations. 
While the methods for locating unstable periodic orbits of high-dimensional 
dynamical systems are well developed,\rf{Visw07b} to the best of 
our knowledge, 
there exists no technique for the unsupervised identification of similarities 
between chaotic trajectory segments and periodic orbits of high-dimensional 
systems. 
In this paper, we shall demonstrate that this can be achieved via 
topological data analysis. 

Topological data analysis is an active field of research with
a continuously growing domain of applications.\rf{ECE2011} 
In a broad sense, topological data analysis methods aim to 
extract significant geometric features of high-dimensional
and/or noisy data sets.
Arguably the most popular tool in this field
is persistent homology,\rf{EH2008,ECE2011} which was 
recently applied to various representations of data 
produced by dynamical systems.
Some examples are 
physical space data obtained from 
biological aggregation models\rf{TZH2015}
and fluid simulations,\rf{KRAMAR2016} 
and time series data from chaotic 
systems.\rf{GaBrMe2016,MyMuKh2019}
Differently from these examples, in the present work, 
we compute persistence in state space
in order to compare the shapes of chaotic trajectory
segments to those of periodic orbits in high-dimensional 
settings.

In this paper, we propose a novel technique for inferring the symbolic 
dynamics of chaotic motion in arbitrary dimensions. 
We name our method ``\perShadow{}'', and illustrate its core ideas on the 
three-dimensional R\"{o}ssler system. 
We then apply the method to the \KS{} partial differential 
equation (PDE) and show that the system's spatiotemporally chaotic dynamics 
can be approximated by a Markov chain
based on four distinct periodic orbits.
The rest of the paper is organized as follows. 
In \refsect{s-Prelim}, we recapitulate the core concepts from the dynamical 
systems theory and topological data analysis, which form the foundations of
\perShadow{}. 
In \refsect{s-ssPersist}, we lay out the steps of \perShadow{} for a generic 
continuous-time dynamical system. 
We demonstrate our method with applications in 
\refsect{s-Numerics}, discuss our results in \refsect{s-discussion} and
conclude in \refsect{s-conclusion}.

\section{Preliminaries }
\label{s-Prelim}

We consider dynamical systems defined by a $D$-dimensional state 
space $\ssp \subset \reals^D$ and a smooth flow map $\flow{\zeit}{\svec}$ 
that maps state vectors $\svec \in \ssp$ as
\beq
	\svec (\zeit) = \flow{\zeit}{\svec (0)} \,, \label{e-Flow}
\eeq
where $\zeit \in \reals^+$ is the time variable. 
Although it is not a general requirement, in the examples we
consider, $\flow{\zeit}{\svec (0)}$ is related to an 
ordinary differential 
equation (ODE) 
\beq
	\dot{\svec} = \vel (\svec) \label{e-ODE}
\eeq
through the relation
\beq
	\flow{\zeit}{\svec (0)} = \svec (0) 
							+ \int_0^\zeit \vel(\svec (\zeit') ) 
							d\zeit' \,, \label{e-FlowIntegral}
\eeq
where $\vel (\svec)$ is called the state space velocity. 
While in the examples worked out here we always use the
	  Euclidean inner product
\beq
	\inprod{\svec^{(i)}}{\svec^{(j)}} = 
	\sum_{k=1}^{D} \svec^{(i)}_{k} \svec^{(j)}_{k} \,, \label{e-inprod}
\eeq
we expect that the topological methods we develop here 
	  do not depend strongly on the particular choice of norm.
In \refeq{e-inprod}, 
subscripts denote vector components and 
superscripts in parentheses denote labels.

\subsection{Symbolic dynamics and shadowing} 
\label{e-SymDyn}

We assume that the state space \ssp{} is coarse-grained into regions 
$\ssp^{(A)}, 
\ssp^{(B)}, 
\ssp^{(C)} \ldots $ 
such that a trajectory $\svec (\zeit)$ for 
$\zeit \in [0, \zeit_F]$
can be associated with an itinerary 
$\symb_1 
 \symb_2 
 \symb_3 \ldots $ with 
$\symb_i \in \{A, B, C, \ldots \}$ 
according to the successive state space regions visited by the trajectory. 
Further, we assume that the system admits periodic orbits 
such that every point $\svec^{(p)}$ on a periodic orbit $p$ satisfies
\beq
\svec^{(p)} = \flow{T_p}{\svec^{(p)}} \label{e-PO}
\eeq
for a nonzero period $T_p$ and its integer multiples.
By definition, a periodic orbit has a cyclic itinerary, such as 
$\overline{\symb_1 
		   \symb_2 
		   \ldots 
		   \symb_n}$, 
where the overline denotes 
infinite repetition. 
In what follows, we use the itinerary 
of a periodic orbit as its label when an itinerary is known. 
Finally, we assume that the first $n$ symbols in the itinerary of 
a trajectory $\svec(\zeit)$ for $\zeit \in [0, \zeit_F]$ 
are the same with that of the 
periodic point $\svec^{(p)}$ if $\svec (0)$ and 
$\svec^{(p)}$ are sufficiently close in 
an appropriately defined state space distance measure.
When a segment of an itinerary of a generic trajectory is the same with 
that of a periodic orbit, we say that ``the trajectory `shadows' the 
periodic orbit''. 
Let us illustrate these concepts with an example. 

\begin{figure}
	\subgraphicw{0.5}{a}{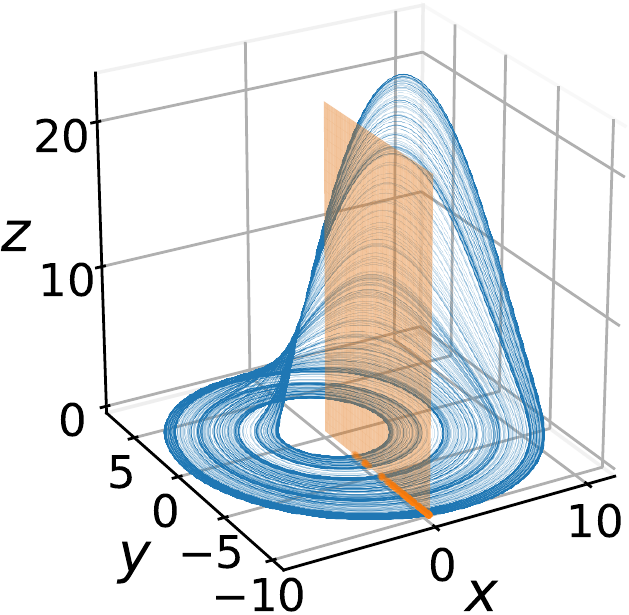}
	\subgraphicw{0.5}{b}{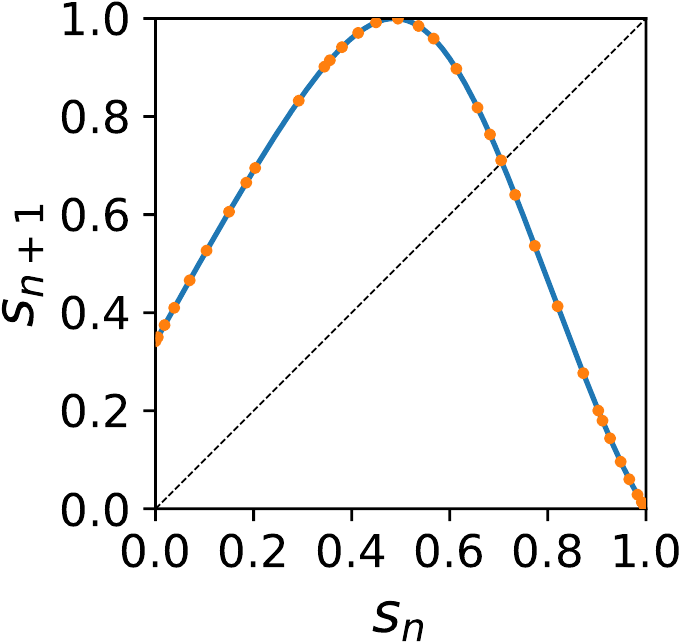}
    \caption{(a) A trajectory (blue) on the R\"{o}ssler attractor 
    		 and its intersections (orange points)
             with the Poincar\'e section \refeq{e-Psect}.
             The Poincar\'e section \refeq{e-Psect}
             is visualized as a transparent surface.
             (b) First-return data (orange) and the  
             Poincar\'e map (blue),
             parameterized by the arc length 
             along the interpolation
             of the intersections in (a). 
             In both (a) and (b), every tenth intersection
             is shown for better visibility.
		\label{f-Rossler}}
\end{figure}

The R\"{o}ssler system is defined by the set 
of ODEs\rf{ross}
 \beq
 \dot{x} = - (y + z) \,,\quad
 \dot{y} = x + 0.2 y \,,\quad  
 \dot{z} = 0.2 + z (x - 5.7) \, . \label{e-Rossler}
 \eeq
The numerical integration of \refeq{e-Rossler} 
reveals a chaotic 
attractor, which we visualize in \reffig{f-Rossler} (a) 
by a trajectory on it.
The temporal length of this trajectory is $\zeit_f = 2000$,
which covers the attractor sufficiently for visualization purposes.
We define a Poincar\'e section $\sspRed$ as the half hyperplane of points 
$\svecRed \in \sspRed$ which satisfy
\beq
\begin{aligned}
    \inprod{\svecRed - \svecRed'}{\eta} &= 0 &
    &\text{and} &
    \inprod{\vel(\svecRed)}{\eta} &> 0 \,, \label{e-Psect}
\end{aligned}
\eeq
where $\svecRed'$ and $\eta$  are called the ``section template'' and the 
``section normal'', respectively.
For the choices of $\svecRed' = (0, -1, 0)$ and $\eta = (1, 0, 0)$ we 
show the Poincar\'e section defined by \refeq{e-Psect} in 
\reffig{f-Rossler} (a) as a transparent surface. 

Let $\svecRed [n]$ be a state vector on the Poincar\'e section \refeq{e-Psect} 
at the discrete time $n$. 
The Poincar\'e map is the discrete-time system 
\beq
	\svecRed [n+1] = \PoincM (\svecRed[n]) 
   				   = \flow{\Delta t_n}{\svecRed} \,,
\eeq
where $\Delta t_n$ is the ``first-return time'', that is, the minimum 
time required for the trajectory of $\svecRed[n]$ to intersect the Poincar\'e 
section \refeq{e-Psect}.
As illustrated by \reffig{f-Rossler} (a), the trajectories 
on the R\"ossler attractor intersect the Poincar\'e section \refeq{e-Psect} 
along what appears to be a one-dimensional curve. 
This suggests the arc length along this curve as a natural 
parametrization for the Poincar\'e map.
We interpolate this curve with cubic splines and use the data to obtain the 
unimodal Poincar\'e return map shown in \reffig{f-Rossler} (b). 

\begin{figure}
    \subgraphicw{0.30}{a}{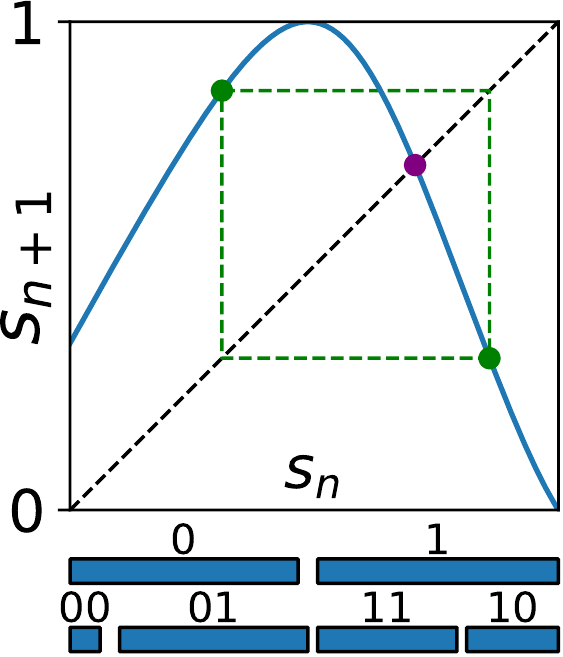}
    \subgraphicw{0.35}{b}{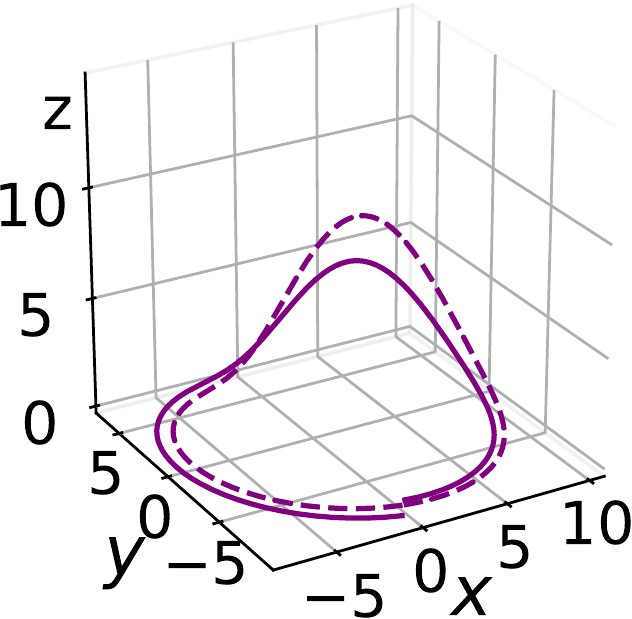}
    \subgraphicw{0.35}{c}{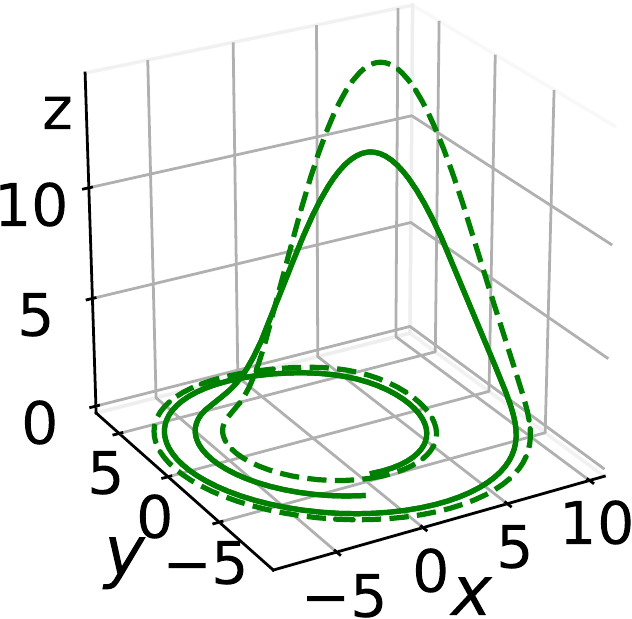}
	\caption{
        (a) Periodic orbits
        $\overline{1}$ (purple point),
        $\overline{01}$ (green points, connected with dashed line 
        segments), and the partitioning of the R\"ossler system's 
        Poincar\'e map.
		(b) Periodic orbit
		$\overline{1}$ (purple, dashed)
		and a shadowing trajectory segment (purple, solid)
		of the R\"ossler system.
		(c) Periodic orbit
		$\overline{01}$ (green, dashed)
        and a shadowing trajectory segment (green, solid)
        of the R\"ossler system.
		\label{f-RosslerMarkov}}
\end{figure}

We are now in position to partition the state space of the R\"ossler 
system into 
regions. 
The return map of \reffig{f-Rossler} (b) has one critical point 
$\svecRed^{(c)} \approx 0.4868$, 
at which the derivative of the Poincar\'e map is zero. 
Let us define regions $\sspRed^{(0)}$ and 
$\sspRed^{(1)}$ as
\bea
	\sspRed^{(0)} &=& 
	\{\svecRed \in \sspRed \,|\, \svecRed < \svecRed^{(c)} \} \,, \\
	\sspRed^{(1)} &=& 
	\{\svecRed \in \sspRed \,|\, \svecRed > \svecRed^{(c)} \} \,.
\eea
With these definitions, we can now assign each trajectory on the R\"ossler 
attractor a binary symbol sequence. 
In particular, we can now enumerate the periodic orbits of the 
R\"ossler system 
with binary numbers. 
\refFig{f-RosslerMarkov} (a) shows the two shortest periodic orbits 
$\overline{1}$ and $\overline{01}$ of the R\"ossler system on the 
Poincar\'e map 
and \reffig{f-RosslerMarkov} (b--c) shows these orbits in the full 
state space. 

It is straightforward to confirm that a
point on the Poincar\'e map 
\reffig{f-RosslerMarkov} (a) that is close to a periodic orbit has
initially 
the same itinerary as that of the periodic orbit. 
We show examples of such ``shadowing'' trajectories
along with the 
periodic ones in 
\reffig{f-RosslerMarkov} (b--c). 
At the bottom of \reffig{f-RosslerMarkov} (a), we show the 
subpartitioning of the unimodal map with respect to the itineraries of the points 
on it.
This partitioning can be confirmed by inspection. 
For example, an initial condition picked from partition $\sspRed^{(10)}$
starts out at partition $\sspRed^{(1)}$ and lands at partition $\sspRed^{(0)}$ after
one iteration of the Poincar\'e map.
Further
iterates of the map would result in finer partitions with 
longer and longer periodic orbits.
For details, we refer the interested reader to 
\refrefs{ASY1997,DasBuch}.

The similarities of the periodic orbits and the shadowing 
trajectory segments in \reffig{f-RosslerMarkov} (b--c) constitute the 
key intuition 
of \perShadow{}. 
In general, it is not possible to reduce the dynamics of a chaotic 
attractor into a unimodal map such as \reffig{f-Rossler} (b). 
However, one can still find periodic orbits and compare the shapes 
of trajectory 
segments to those of the periodic orbits. 
Our next step is to introduce persistent homology which we utilize 
for this 
purpose.

\subsection{Persistent homology} 
\label{s-persist}

Persistent homology is a mathematical framework
for extracting significant shapes in a data set.
In this section, we illustrate the persistent homology concepts 
that we incorporate in our method through an example, 
while trying 
to avoid the technical language as much as possible. 
For in-depth introductions, 
we refer the reader to the survey\rf{EH2008} and 
the textbook\rf{EH2010} by Edelsbrunner and Harer and the 
``roadmap''\rf{Otter2017} by Otter \etal{}

\begin{figure*}
	\subgraphicw{0.127}{a}{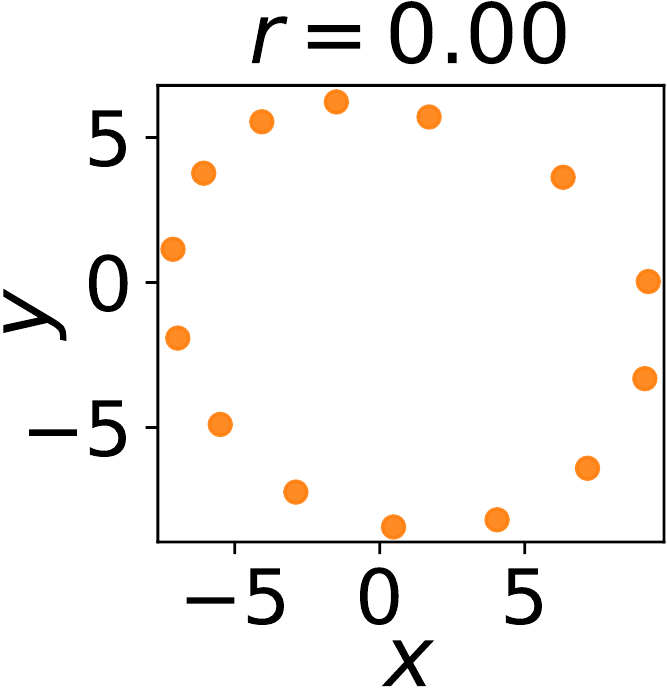}\,
	\subgraphicw{0.127}{b}{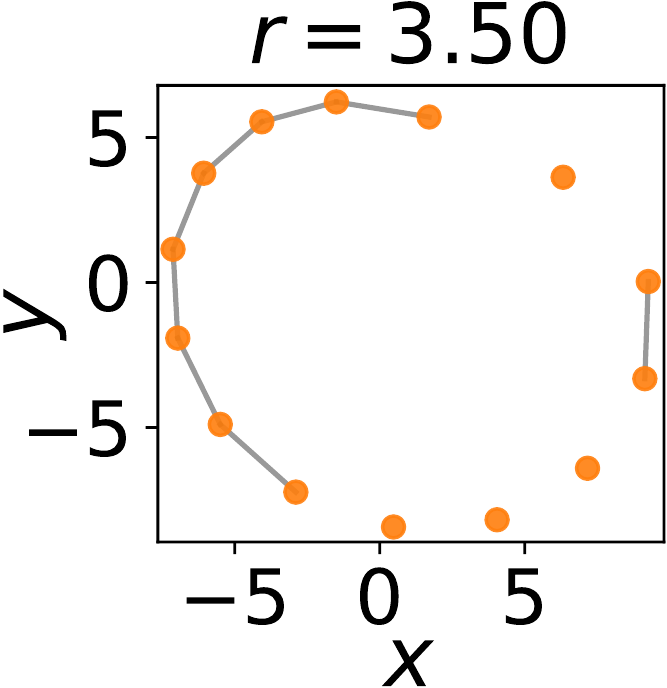}\,
	\subgraphicw{0.127}{c}{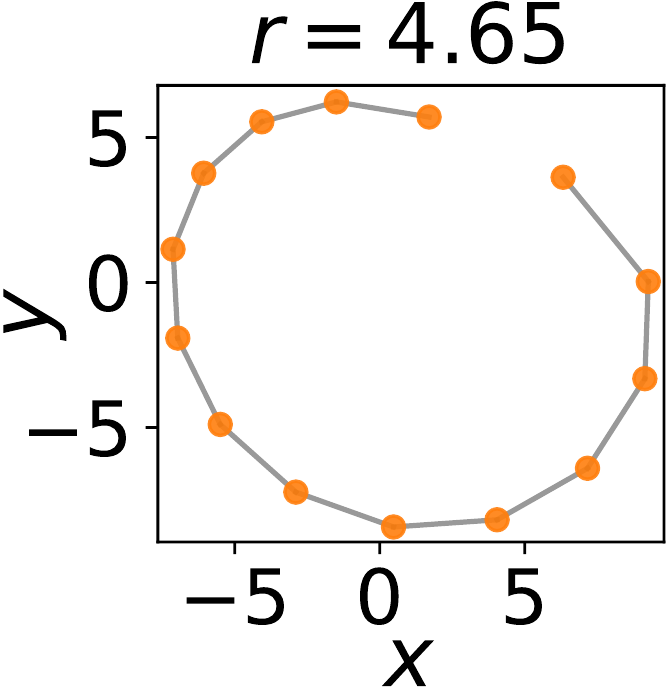}\,
	\subgraphicw{0.127}{d}{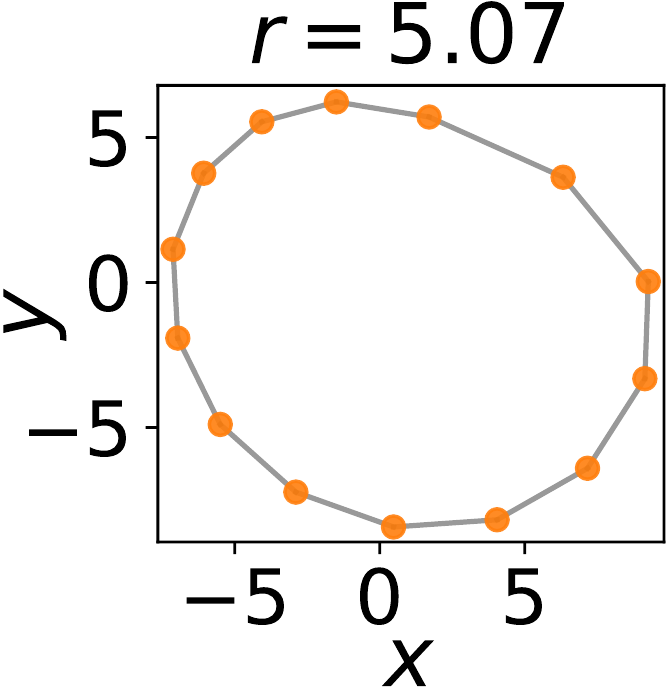}\,
	\subgraphicw{0.127}{e}{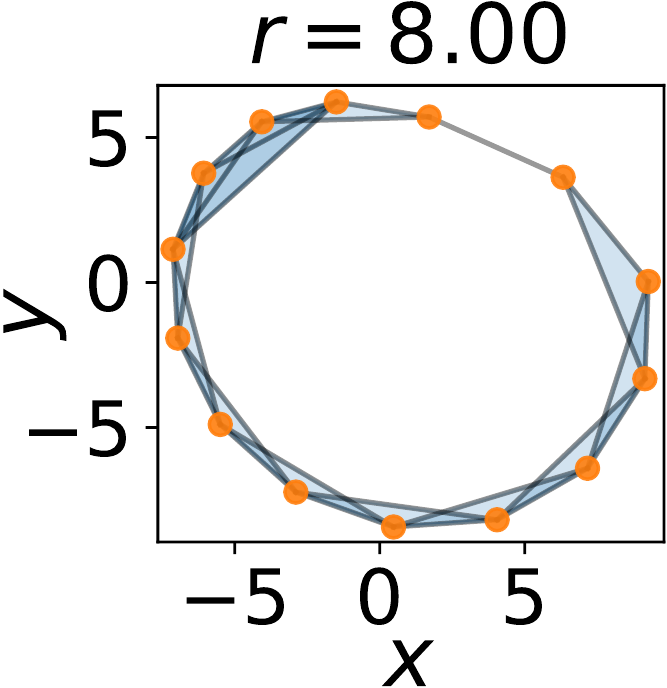}\,
	\subgraphicw{0.127}{f}{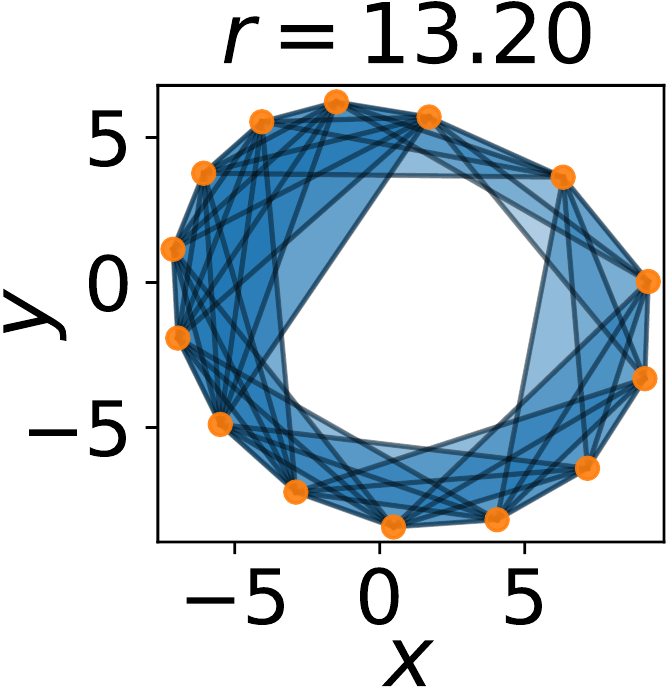}\,
	\subgraphicw{0.127}{g}{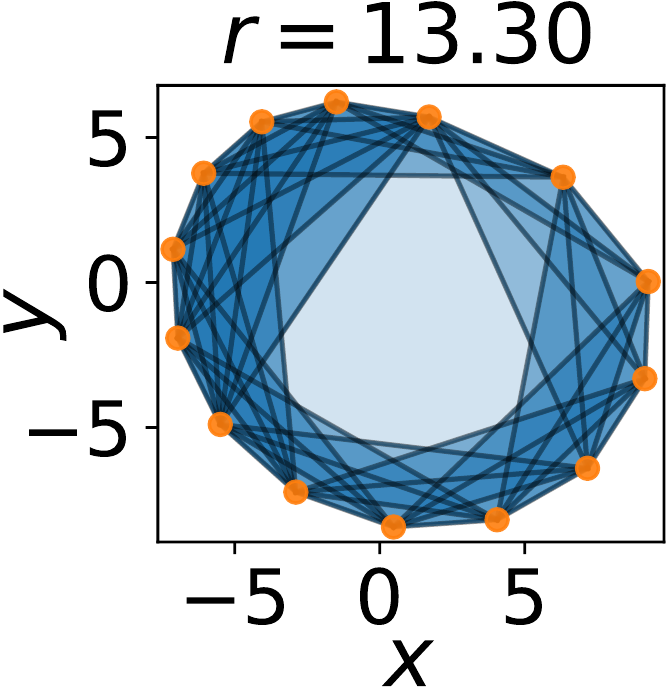}\\
	\subgraphicw{1}{h}{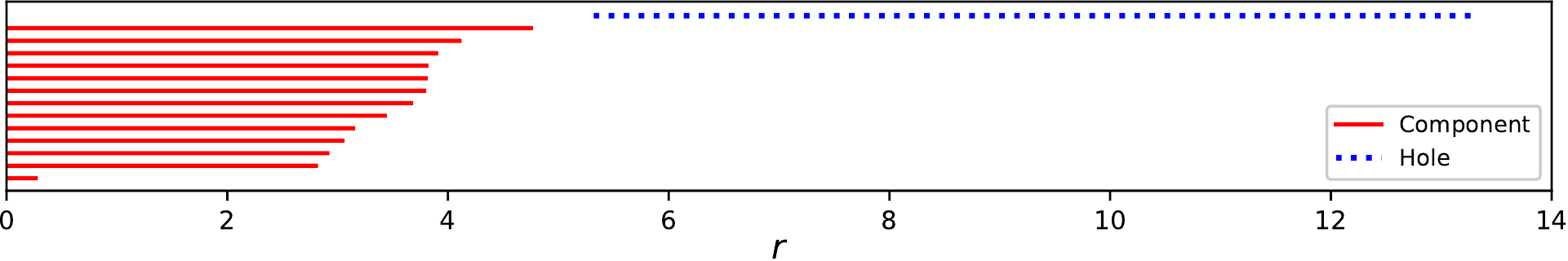} 
	\caption{
        (a--g) 
        Visualizations of the sequence of sets 
        $\DataSet_{\resol}$ corresponding to a data set 
        sampled from the periodic orbit $\overline{1}$ of the R\"ossler 
        system.
        The parameter $\resol$ is a distance threshold that sets the 
       	connectivity of the points in the data set: Two points
        that are closer than $\resol$ are connected with a line and 
        three points
        that are pairwise closer than $\resol$ form a triangle.
        In all figures, the initial data points are shown orange, edges 
        connecting them are drawn as line segments, triangles are 
        visualized as transparent blue fillings,
        and the values of $\resol$
        are noted on top.
    	(h) Barcode diagram showing the birth and death of components 
    	(red, solid) and 
   		holes (blue, dotted) as $\resol$ increases.
   		Each bar spans an interval that begins at the $r$
		value at which the respective component or hole is born
   		and ends at the $r$ value at which it dies.
   	\label{f-RosslerPH}}
\end{figure*}

In our applications, we consider a data set
\beq 
\DataSet = \{\svecproj^{(1)}, \svecproj^{(2)}, \ldots, \svecproj^{(N)} \}
\label{e-DataSet}
\eeq
that is composed of projections $\svecproj^{(i)} = \mathbf{P} \svec^{(i)}$
	  of state vectors $\svec^{(i)}$ sampled from a trajectory of 
	  a dynamical system. 
	  $\mathbf{P}$ is a projection operator  
	  specific to the application.
\refFig{f-RosslerPH} (a) shows an example of such a data set from 
the R\"ossler system as a projection onto the $(x,y)$-plane.
These points are
sampled from 
the periodic orbit $\overline{1}$ (\reffig{f-RosslerMarkov}) of the R\"ossler
system with a constant time step of $\zeit_s=0.45$. 

For the analysis to follow, we need a distance function 
	  for the projected data set, which we define as
\beq
d(\svecproj^{(i)}, \svecproj^{(j)}) 
= \inprod{\svecproj^{(i)} - \svecproj^{(j)}}{
	\svecproj^{(i)} - \svecproj^{(j)}}^{1/2} 
\, .
\label{e-dataMetric}
\eeq

Let $\resol \ge 0$ be the ``resolution'' (a distance parameter), 
$\DataSet_{\resol}$ denote a continuous 
sequence of sets of subsets of $\DataSet$ parameterized by $\resol$, and
$\DataSet_0 = \{ \{ \svecproj^{(1)} \}, \{ \svecproj^{(2)} \}, 
			       \ldots, \{\svecproj^{(N)}\} \}$. 
The sets
$\DataSet_{\resol}$ are 
formed by the union of $\DataSet_0$ with all edges 
$\{\svecproj^{(i)}, \svecproj^{(j)} \}$ such that 
$d(\svecproj^{(i)}, \svecproj^{(j)}) \le \resol$ and all triangles 
$\{\svecproj^{(i)}, \svecproj^{(j)}, \svecproj^{(k)} \}$ such that all
pairwise distances
$d(\svecproj^{(i,j,k)}, \svecproj^{(i,j,k)}) \le \resol$. 
In general, this sequence is extended to include tetrahedrons 
		and higher-dimensional generalizations. 
		However, we stop at triangles since this is sufficient 
		for our applications. 
We show visual representations of
$\DataSet_{\resol}$ for different values of $\resol$ in 
\reffig{f-RosslerPH} (a--g).

\begin{figure}
	\subgraphicw{0.5}{a}{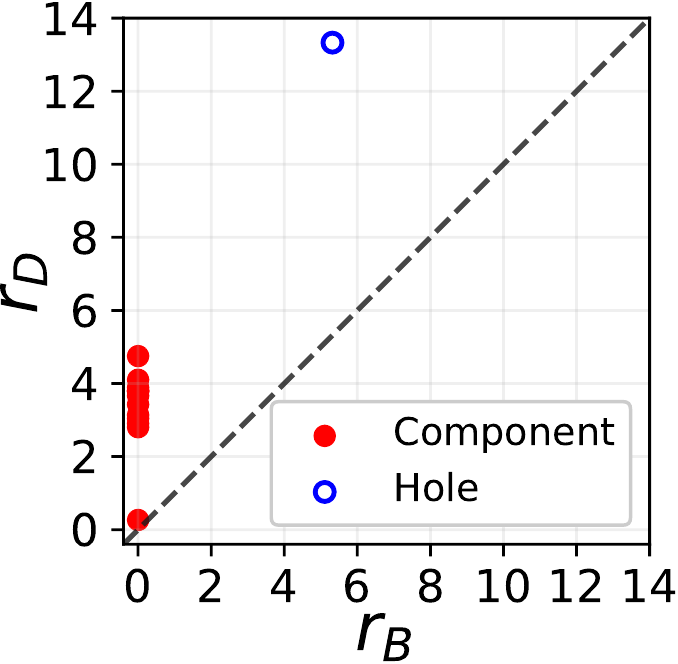}
	\subgraphicw{0.5}{b}{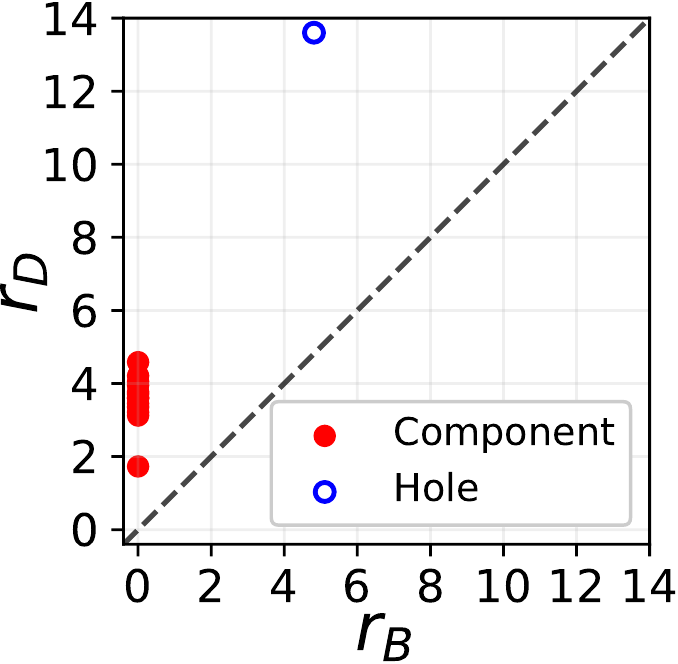}	
	\caption{
		(a) Persistence diagram corresponding to a data set 
		sampled from the periodic orbit $\overline{1}$ of the R\"ossler 
		system.
        (b) Persistence diagram obtained from a R\"ossler system trajectory 
        which shadows the periodic orbit $\overline{1}$ as shown 
        in \reffig{f-RosslerMarkov} (b).
        The birth and death coordinates 
        $(\resol_B, \resol_D)$ of the components and holes are 
        marked red/solid and blue/hollow respectively.
		\label{f-RosslerPHdiagrams}}
\end{figure}

As we vary $\resol$ from $0$ to $\infty$, we keep track of 
the number 
of components
and holes in $\DataSet_{\resol}$.
By a ``component'', we to refer to an individual 
point or a set of points and the edges that
connect them and all triangles that fill the 
				   space in between. 
A ``hole'' is formed when a component is in the form of a 
	  loop with not-necessarily-distinct inner and outer 
  	  boundaries.			   
For example, in \reffig{f-RosslerPH} (a) we have $14$ components, whereas in 
\reffig{f-RosslerPH} (b) we have $6$ and in 
\reffig{f-RosslerPH} (c) we have $1$.
In \reffig{f-RosslerPH} (d), the single component of 
$\DataSet_{\resol = 5.07}$ 
forms a loop with a hole. 
As we further increase $\resol$, triangles begin to form 
(\reffig{f-RosslerPH} (e--f)), and finally, the hole is completely filled 
with triangles at $\resol = 13.30$ (\reffig{f-RosslerPH} (g)).
This sequence of appearances, called ``birth'',
and 
disappearances, called ``death'', of shapes can 
be encoded 
into diagrams such as the one in
\refFig{f-RosslerPH} (h). 
\refFig{f-RosslerPH} (h) is called a ``barcode diagram'', where components 
and holes
are represented by bars that span the interval of $\resol$ 
for which the respective object can be observed. 
Another graphical representation of the same information is the so-called 
``persistence diagram'', on which the birth and death coordinates 
$(\resol_B, \resol_D)$ of components and holes are marked as shown in 
\reffig{f-RosslerPHdiagrams}.
We would like to note here that when two points are connected
	by an edge, which of the two points dies is ambiguous. 
	This ambiguity, however, does not affect the barcode 
	and persistence
	diagrams since both points appear at $\resol = 0$.

Given a data set $\DataSet$, the object that is of interest to us is
the 
associated persistence diagram $\PerD (\DataSet)$. 
In general, the elements that are further away from the diagonal of a 
persistence diagram are said to be the more significant features of 
the data set, 
since they live for a longer range of resolutions.\rf{Carlsson2009}
However, depending on the problem and what the resolution $r$ represents,
features of interest may appear as short-lived elements as 
well.\rf{FengPorter2019,StolzEtAl2017}

An important property of persistence diagrams is
their stability:
If the samples in the data set $\DataSet$ are slightly perturbed, 
then the associated persistence diagram
changes only slightly.
We illustrate this in \reffig{f-RosslerPHdiagrams} where we
show the
persistence diagram 
associated with a data set sampled from
the periodic orbit $\overline{1}$ (\reffig{f-RosslerMarkov} (b), dashed)
of the Rössler system next to the persistence diagram 
of a data set sampled from
a trajectory that shadows it (\reffig{f-RosslerMarkov} (b), solid).
A proof of the stability of 
persistence diagrams can be found in \refref{CEH2007}.

We quantify the similarity of two persistence diagrams by defining a 
distance between them. 
Let us first define the set of diagonal elements
$\Delta = \{ (\resol_B, \resol_D) \in [0, \infty) \times [0, \infty)\, 
|\, \resol_B = \resol_D \}$.
These correspond to the ``trivial'' persistence diagram 
elements that are born 
and die at the same $\resol$ value. 
We can also denote the components and holes on a persistence diagram by the 
multisets 
 \bea
 	\PerD_{i} &=& \{(\resol_B, \resol_D)_{i, 1}, 
 			     (\resol_B, \resol_D)_{i, 2}, 
 			     \ldots 
 			     (\resol_B, \resol_D)_{i, E_i}\} \continue 
 		   && \, \cup \Delta \cup \Delta \cup \Delta \ldots
 		   \,,\quad i \in \{0,1\} \,, \label{e-PerDset}
 \eea
where $i = 0$ corresponds to the components, $i = 1$ corresponds to the 
holes, and $E_i$ is the number of respective elements in a diagram.
We included the trivial sets with infinite multiplicity into the persistence 
diagrams for a reason which will be apparent soon.
We are now in position to define a metric between the persistence 
diagrams $\PerD_{i}^{(n)}$ and $\PerD_{i}^{(m)}$. 
Let $\phi: \PerD_{i}^{(n)} \rightarrow \PerD_{i}^{(m)}$ be a bijection that 
pairs each element of $\PerD_{i}^{(n)}$ with exactly one element of 
$\PerD_{i}^{(m)}$. 
We define the $p^\text{th}$ Wasserstein distance between 
$\PerD_{i}^{(n)}$ 
and $\PerD_{i}^{(m)}$ as
\beq
	W_{p} (\PerD_{i}^{(n)}, \PerD_{i}^{(m)}) 
	= \inf_{\phi} \left[ 
	 \sum_{\PDelm \in \PerD_{i}^{(n)}} 
	 \|\PDelm - \phi(\PDelm)\|_q^p \right]^{1/p}
	 \,, \label{e-Wasserstein}
\eeq
where $p \in [1, \infty]$, $q \in [1, \infty]$, and $\|\,\|_{q}$ denotes 
the $L_q$-norm in $\reals^2$.
When $p=1$,
the Wasserstein distance \refeq{e-Wasserstein} can be understood as 
the smallest 
possible sum of the lengths of the line segments that can be drawn from the 
elements of $\PerD_{i}^{(n)}$ to those of $\PerD_{i}^{(m)}$. 
The addition of diagonal elements to the persistence diagrams makes it 
possible to compare
different diagrams with possibly different number of 
nontrivial elements by allowing matching the nontrivial elements of 
one diagram 
to the diagonal of the other.
Setting $p > 1$ emphasizes the contributions from 
the elements that are further away from the diagonal in comparison 
to others, \ie\ the ones that are more persistent against the changes in
$\resol$.

This concludes our overview of the persistent homology concepts that we
incorporate into \perShadow{}. 
There are various algorithms and implementations for computing persistence 
diagrams and the Wasserstein distances \refeq{e-Wasserstein} 
between them, which 
are not in 
the scope of the present work. 
For a review, we refer the interested reader to 
\refref{Otter2017}.
In the applications that we present in \refsect{s-Numerics}, we 
utilize
the programs \texttt{Ripser}\rf{ripser} for the
computation of persistence diagrams and \texttt{Hera}\rf{hera,KMN2016} for the 
computation of the Wasserstein distance.

\section{\PerShadow{}}
\label{s-ssPersist}

In this section, we list the basic steps of \perShadow{} 
for capturing the symbolic dynamics of high-dimensional chaotic flows. 
Since
the primary applications we have in mind are discretizations
of nonlinear 
PDEs such as the Navier--Stokes equations, 
our presentation below is given for such systems.

\subsection{Symmetry reduction}
\label{s-symred}

Nonlinear PDEs are usually equivariant under a certain set of symmetries 
such as 
translations, rotations, and reflections. 
These symmetries tend to obscure the dynamics by increasing the data volume 
since each solution has a set of symmetry copies that are also solutions. 
Furthermore, systems with continuous symmetries tend to have 
higher-dimensional 
invariant solutions such as relative periodic orbits,\rf{ChossLaut00,DasBuch} 
which are periodic orbits up to continuous symmetry transformations. 

For \perShadow{}, we assume that 
$(\ssp, \timeflow)$ is a 
symmetry-reduced realization of the dynamical system under consideration. 
In other words, before we begin our analysis, we carry out a symmetry-reducing
coordinate transformation, which maps each symmetry-equivalent solution of the 
system to a single representative $\svec \in \ssp$. 
This, in general, can be a nontrivial task. 
However, there has been considerable development in recent years following the 
introduction of the ``first Fourier mode slice'',\rf{BudCvi14} which
is a straightforward method for reducing the $SO(2)$ symmetry due to 
translation equivariance and periodic boundary conditions.
Since its introduction, this method was adapted to the two-dimensional 
Kolmogorov 
flow,\rf{Faraz15} three-dimensional pipe flow,\rf{BudHof17,BudHof18} 
one-dimensional Korteweg--de Vries equation,\rf{MowSap2018}
and pilot-wave hydrodynamics.\rf{BudFle19} 
For a pedagogical introduction to the first Fourier mode slice, we refer the 
reader to \refref{BuBoCvSi14}. 
The reduction of discrete symmetries can also be nontrivial. 
The only discrete symmetry reduction method for high-dimensional systems 
in the 
literature known to us is the invariant polynomials for reflection-type 
symmetries.\rf{BudCvi15}
We present the symmetry reduction of the
\KS{} system in \refappe{a-symred}.

\subsection{Base set of periodic orbits}

We search for a base set of periodic orbits 
$\po = \{\po_1, \po_2, \ldots, \po_M \}$, with which we
attempt to 
approximate chaotic dynamics. 
Generically, this set of periodic orbits can be found via recurrence-based 
searches\rf{SCD07,Visw07b,CviGib10,LucKer14,WFSBC15} or following 
bifurcations\rf{Christiansen97,KreEck12} and unstable manifolds of known 
solutions.\rf{lanCvit07,BudCvi15}
While there exist variational,\rf{lanCvit07} Levenberg--Marquardt 
search-based,\rf{SCD07} and possibly various other optimization methods for 
numerically locating unstable periodic orbits, the current 
community standard for very-high-dimensional flows is the 
Newton--Krylov--hookstep 
method of Viswanath.\rf{Visw07b}

\subsection{Local persistence of periodic orbits}
\label{e-persistPO}

We sample the states $\svec^{(\po_i)} (t)$ on each periodic orbit 
$\po_i$ with a constant sampling time 
$\zeit_s$ and construct
local projection bases
$\{e_1^{(\po_i)},\, e_2^{(\po_i)},\, \ldots,\, e_{N_i}^{(\po_i)}\}$ 
with the origins
$O^{(\po_i)}$ that locally capture the data points 
$\{\svec^{(\po_i)} (0),\, \svec^{(\po_i)} (t_s),\, \ldots,\, 
		   \svec^{(\po_i)}((N_i - 1) t_s) \}$
 of $\po_i$. 
This can be achieved by a standard method such as principal component 
analysis (PCA).\rf{Jolliffe2002}
Note that with a fixed sampling time, each periodic orbit $\po_i$ has 
a different number of samples $N_i$. 
Finally, we generate a catalog of persistence diagrams 
$\PerD^{(\po_1)}, \PerD^{(\po_2)}, \ldots, \PerD^{(\po_M)}$ from the
local projections of the periodic orbit samples. 

\subsection{Local persistence of chaotic trajectory segments}
\label{e-persistChaotic}

Consider the data set 
\beq
\DataSet^{(i)} (\zeit) 
= \{\svecproj (\zeit), 
	\svecproj (\zeit+\zeit_s),\, 
	\svecproj (\zeit+2\zeit_s),\,\ldots,\, 
	\svecproj (\zeit+(N_{i}-1)\zeit_s)\} \label{e-DataSetChaos}
\eeq
with $N_{i}$ elements that are sampled from a chaotic trajectory starting 
at time $\zeit$ and projected onto the local bases of the 
			  $i^{th}$ periodic orbit. 
Let $\PerD^{(i)}(\zeit)$ be the persistence diagram
obtained from 
this data set. 
We define the \shadowDist{} of a chaotic trajectory segment to the periodic 
orbit $\po_i$ at time $\zeit$
as the weighted sum 
\bea
	\distShadow^{(i)} (t) 
	&=& w_0 W_p (\PerD_{0}^{(i)}(\zeit),\, \PerD_{0}^{(\po_i)}) \continue
	&+& w_1 W_p (\PerD_{1}^{(i)}(\zeit),\, \PerD_{1}^{(\po_i)}) \, .
	\label{e-distShadow}
\eea
The adjustable weights $w_0$ and $w_1$ in \refeq{e-distShadow} 
control the respective contributions of the components and the
holes to the shadowing distance.
In order to identify a chaotic trajectory's transient visits to the 
neighborhoods of the periodic orbits, we measure its \shadowDist{}
from the base set of periodic orbits. 
The set of \shadowDist{}s is the final output of \perShadow{}. 
As we shall see in our applications, 
the shadowing distance of a chaotic trajectory segment 
to a periodic orbit becomes small when the trajectory 
segment has the same itinerary as the periodic orbit.

\section{Numerical demonstrations}
\label{s-Numerics}

In this section, we present two applications of \perShadow{}. 
We begin with a controlled
numerical experiment on the R\"ossler 
system.

\subsection{R\"ossler system}
\label{s-Roessler}

Since the R\"ossler system has no symmetries, 
we do not need a symmetry reduction, 
thus we begin our analysis by choosing 
$\{\po_1 = \overline{1}, \po_2 = \overline{01}\}$ 
(\reffig{f-RosslerMarkov})
as our base set.
We also do not need local projection bases 
for these periodic orbits. Since the R\"ossler
system is three-dimensional, we can carry out 
the calculation to follow in the full state space.

\begin{figure}
	\graphicw{1}{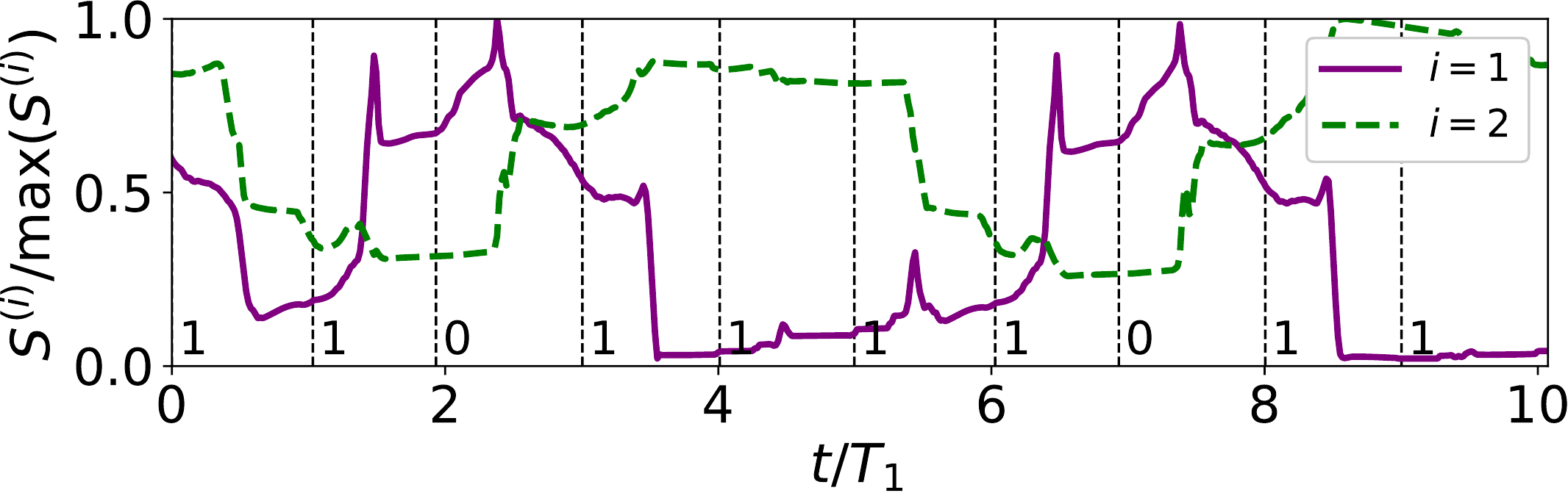}
	\caption{
        Time series of the \shadowDist{}s $\distShadow^{(1)}$  
        and $\distShadow^{(2)}$
        between 
        trajectory segments
        on 
        the Rössler attractor and 
        the periodic orbits
        $\po_1 = \overline{1}$ 
        and  $\po_2 = \overline{01}$.
		The time axis is in units of the period of the orbit
		$\overline{1}$.
        Symbol sequences printed at the bottom 
        correspond to the windows marked by the vertical 
    	dashed lines and 
        are read off from the Poincar\'e section.
		\label{f-RosslerDistShadow}}
\end{figure}

We sample a chaotic trajectory
and the 
periodic orbits $\overline{1}$ and $\overline{01}$
with a constant sampling time of $t_s = 0.1$
and compute the \shadowDist{}s \refeq{e-distShadow}
of the chaotic trajectory from the periodic orbits using 
unit weights \(w_0=w_1=1\) 
and the Wasserstein 
distance \refeq{e-Wasserstein} with \(p=q=2\).
\refFig{f-RosslerDistShadow}
shows the \shadowDist{}s of a 
chaotic trajectory of 
the R\"ossler system from the periodic orbits $\overline{1}$ and 
$\overline{01}$.
We normalized the shadowing distances 
in \reffig{f-RosslerDistShadow} by their respective
maxima so that each time series takes values in the
interval $[0,1]$.
The symbols 
printed at the bottom, separated by vertical dashed 
line segments,
form the itinerary of the chaotic trajectory and
are read off from the Poincar\'e section.
It is clear from \reffig{f-RosslerDistShadow} that the distance of the chaotic 
trajectory to the periodic orbit $\overline{1}$ has a dip when the 
chaotic trajectory's itinerary has a $1$. 
Similarly, the distance to the periodic orbit $\overline{01}$ 
has a dip when the itinerary has a symbol sequence $01$ or $10$. 
These drops in the \shadowDist{} can be easily detected 
using a threshold 
and thus \perShadow{} can indeed be used for inferring symbolic dynamics.

\subsection{\KS{} system}
\label{s-KS}

The \KSe{} was originally proposed to model the phase dynamics 
of reaction-diffusion systems\rf{ku} and instabilities of flame 
fronts.\rf{siv} Owing to its computational simplicity, nowadays the
\KS{} system is frequently chosen as the testing ground for methods to 
study high-dimensional chaos and turbulence.\rf{Holmes96,SCD07,BudCvi15,
	Goluskin_2019,Pathaketal2018,Pathaketal2017} 
In $(1+1)$ dimensions, 
the \KSe{} reads
\beq
u_\zeit = -u\,u_x
-u_{xx}-u_{xxxx} \,,
\label{e-ks}
\eeq
where $x \in [-L/2, L/2)$ and $\zeit \in [0, \infty)$ 
denote the space and time coordinates respectively and
the subscripts imply partial derivatives. We interpret the scalar field 
$u(x, \zeit)$ as the flame front velocity and assume the
periodic boundary condition $u(x, \zeit) = u(x + L, \zeit)$. The domain 
length $L$ is the sole control parameter of the \KS{} system, whose 
dynamics become chaotic when $L$ is large enough.\rf{SCD07,BudCvi15}

The \KSe{} \refeq{e-ks} is equivariant under continuous translations
\beq
	\LieEl_x (\delta x) u(x, \zeit) = u(x - \delta x, \zeit)\,,
	\label{e-gx}
\eeq 
where $\delta x \in [0, L)$, and the reflection
\beq
 	\sigma u(x, \zeit) = -u(-x, \zeit) \,. \label{e-sigma}
\eeq
As a consequence of the symmetries \refeq{e-gx} and \refeq{e-sigma},
the \KS{} system has relative periodic orbits, which satisfy 
\beq
	u_p = \LieEl \flowKS{T_p}{u_p} \,, \label{e-rpo}
\eeq
where $\LieEl \in \{\LieEl_x (\delta x_p), \sigma \}$\,,
$\delta x_p \in [0, L)$, and 
$\flowKS{\zeit}{u}$ is 
the flow map induced by the time evolution under \refeq{e-ks}. 
As we argued in \refsect{s-symred}, before the 
persistence analysis, we must obtain a symmetry-reduced 
representation for the \KS{} system. This problem was addressed 
in \refref{BudCvi15}, which combined the \fFslice{} method of
\refref{BudCvi14} with an invariant-polynomial method to obtain
a fully symmetry-reduced representation of the \KS{} state space.
Here, we follow a slightly different approach that does 
not introduce any new technique, therefore, we leave the 
details of this to \refappe{a-symred} and assume that we have a 
symmetry-reducing transformation $\svecRed = \SymRed(u)$ for all 
$u(x, \zeit)$ of interest, such that
\beq
	\svecRed = \SymRed(u) = \SymRed(\LieEl u) \,, \label{e-symred}
\eeq
where $\LieEl \in \{\LieEl_x (\delta x), \sigma \}$ and 
$\delta x \in [0, L)$.
Once we obtain the symmetry-reducing 
transformation \refeq{e-symred}, the symmetry-reduced flow
is obtained straightforwardly as
\beq
	\svecRed(\zeit) = \flowRed{\zeit}{\svecRed(0)} 
				 = \SymRed(\flowKS{\zeit}{\SymRed^{-1}(\svecRed(0)) }) \, .
	\label{e-flowRed}
\eeq
Note that the inverse transformation 
$u = \SymRed^{-1}(\svecRed)$ cannot be unique, since the 
symmetry reduction
\refeq{e-symred} maps all symmetry-equivalent states to
one. However, this nonuniqueness makes no difference in the
symmetry-reduced flow \refeq{e-flowRed}, thus, any one of the 
available symmetry-equivalent inverses can be taken.

\begin{figure}[h]
	\graphicw{1}{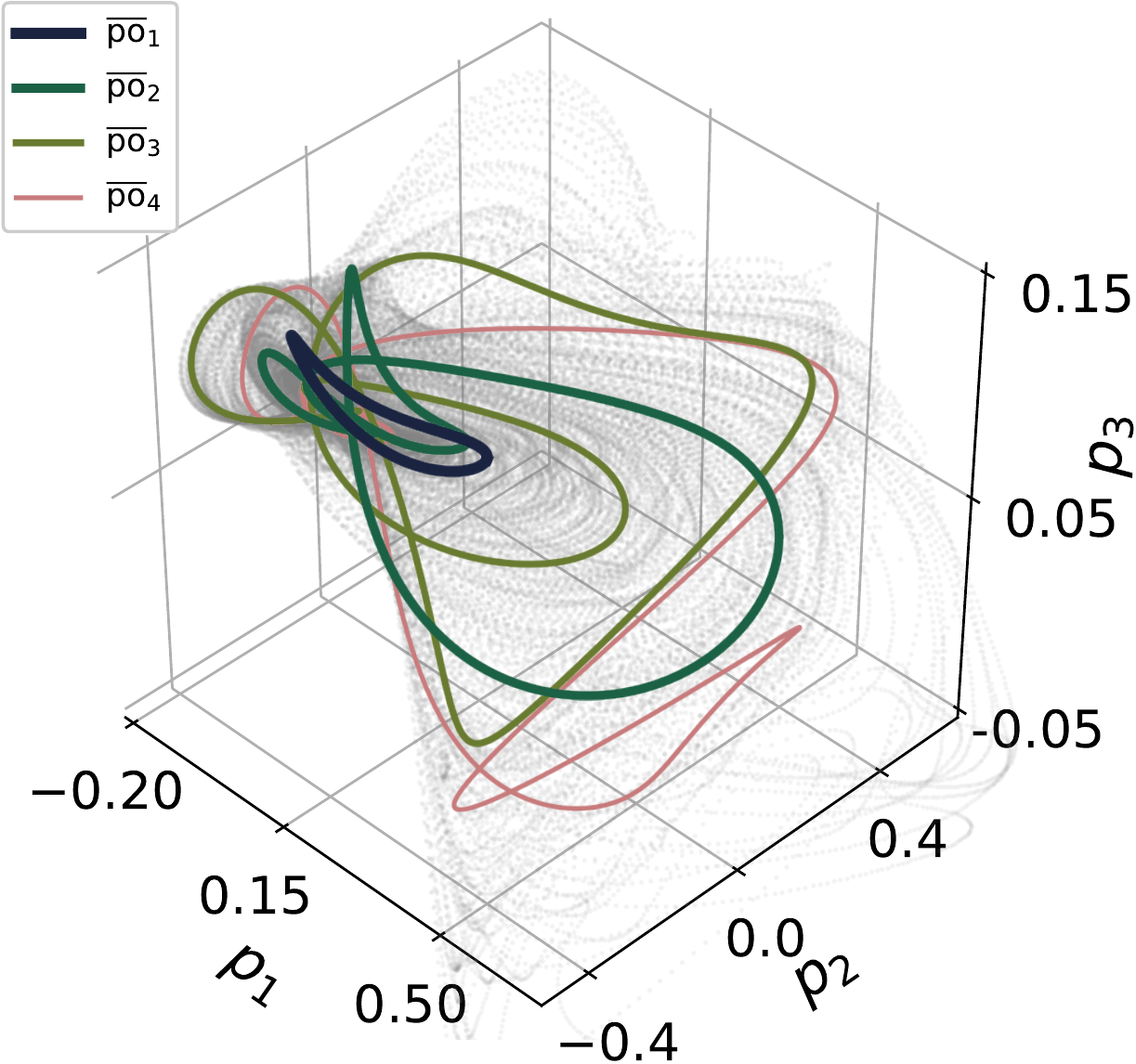}
	\caption{
		A chaotic trajectory (gray dots) and four periodic orbits 
		(different colors/thickness)
		of the \KS{} system projected onto three 
		leading PCA directions obtained from a chaotic data set.
		\label{f-ksAttrGlobal}}
\end{figure}

After the symmetry reduction 
\refeq{e-symred}, by definition, the relative periodic orbits
\refeq{e-rpo} become periodic orbits \refeq{e-PO}.
By numerically following the unstable manifolds of relative 
periodic orbits, \refref{BudCvi15} presented evidence that the 
chaotic dynamics of the \KS{} system at $L = 21.7$ take place in the 
vicinity of four relative periodic solutions all of which are 
unstable. We show
these orbits along with a chaotic trajectory in 
\reffig{f-ksAttrGlobal} as a PCA projection.
The projection
bases were obtained as the first three principal components 
corresponding to a chaotic trajectory 
of temporal length $\zeit_f = 10^5$,
sampled at the sampling time $t_s = 10$. 
Four periodic orbits 
\beq
	\po = \{\po_1, \po_2, \po_3, \po_4 \} \label{e-baseset}
\eeq
with periods 
$T_1 = 10.11,\, T_2 = 32.37,\, T_3 = 36.70,\, T_4 = 36.08$, 
which we show
in 
\reffig{f-ksAttrGlobal}, form the base set for the 
\perShadow{} of the \KS{} system.
In order to confirm that our chaotic data set is long enough
to cover the attractor of the system sufficiently, 
we reproduced \reffig{f-ksAttrGlobal} with random initial conditions
and found the resulting projections to be practically 
indistinguishable. 
We generated these initial conditions by populating 
the Fourier coefficients of $u(x, 0)$ with 
random numbers drawn from the standard normal distribution.

\begin{figure*}
	\subgraphicw{1.0}{a}{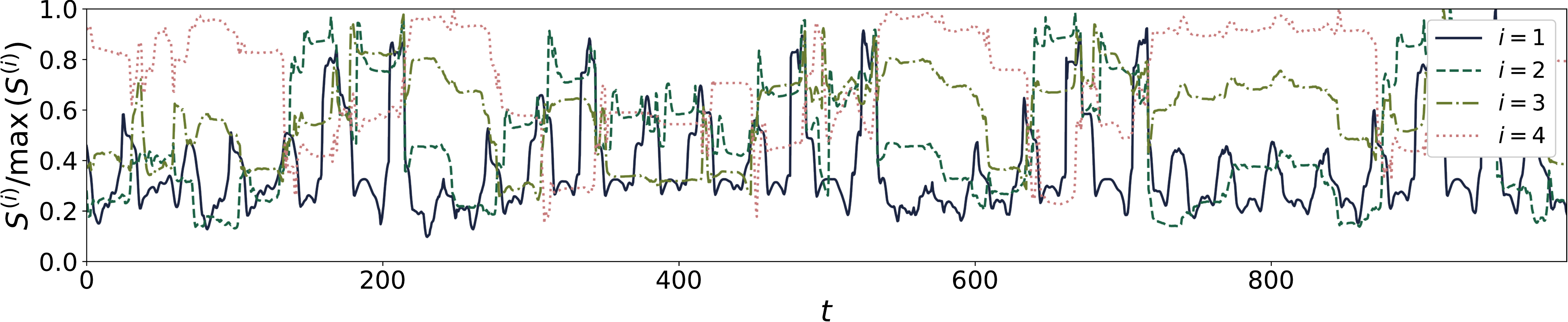} \\
	\subgraphicw{1.0}{b}{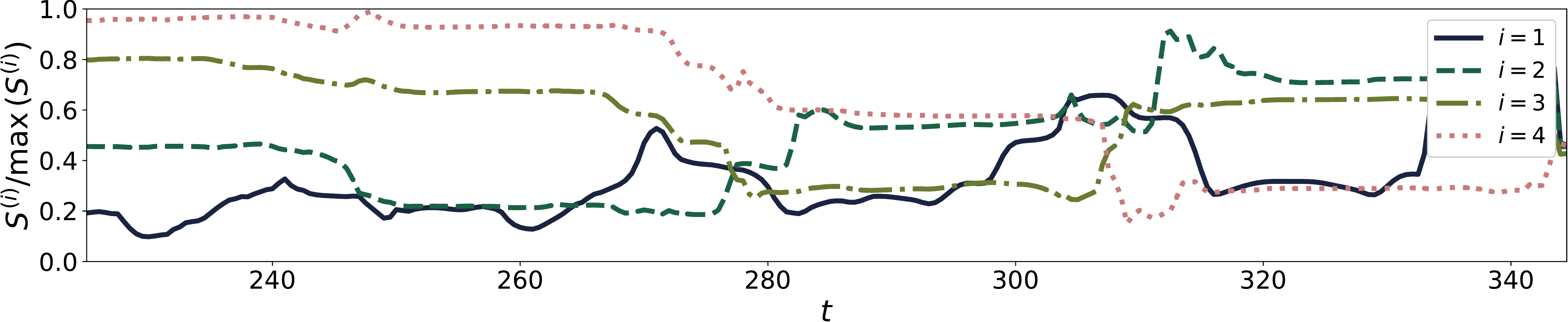} \\
	\subgraphicw{0.231}{c}{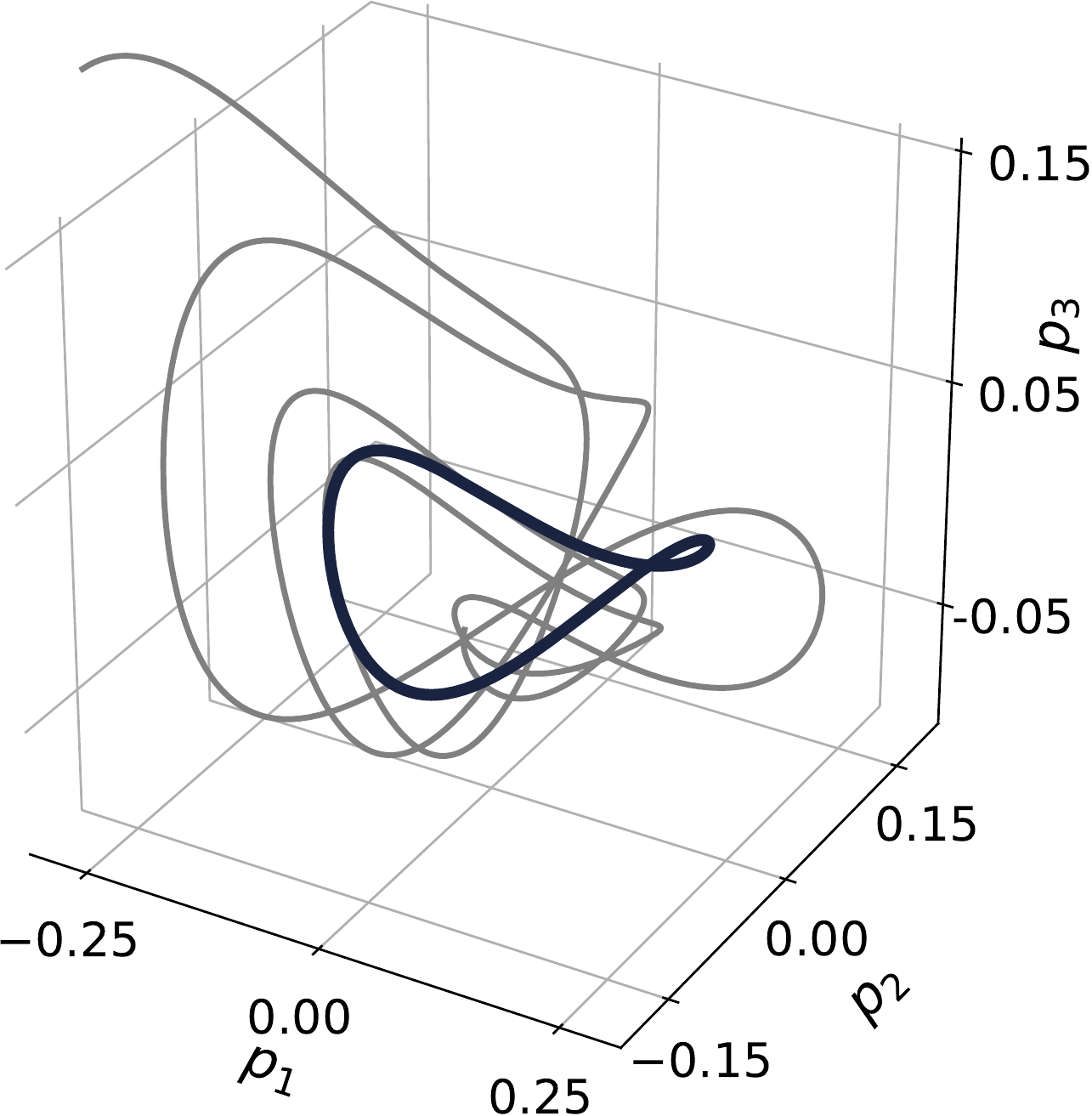}\,
	\subgraphicw{0.23}{d}{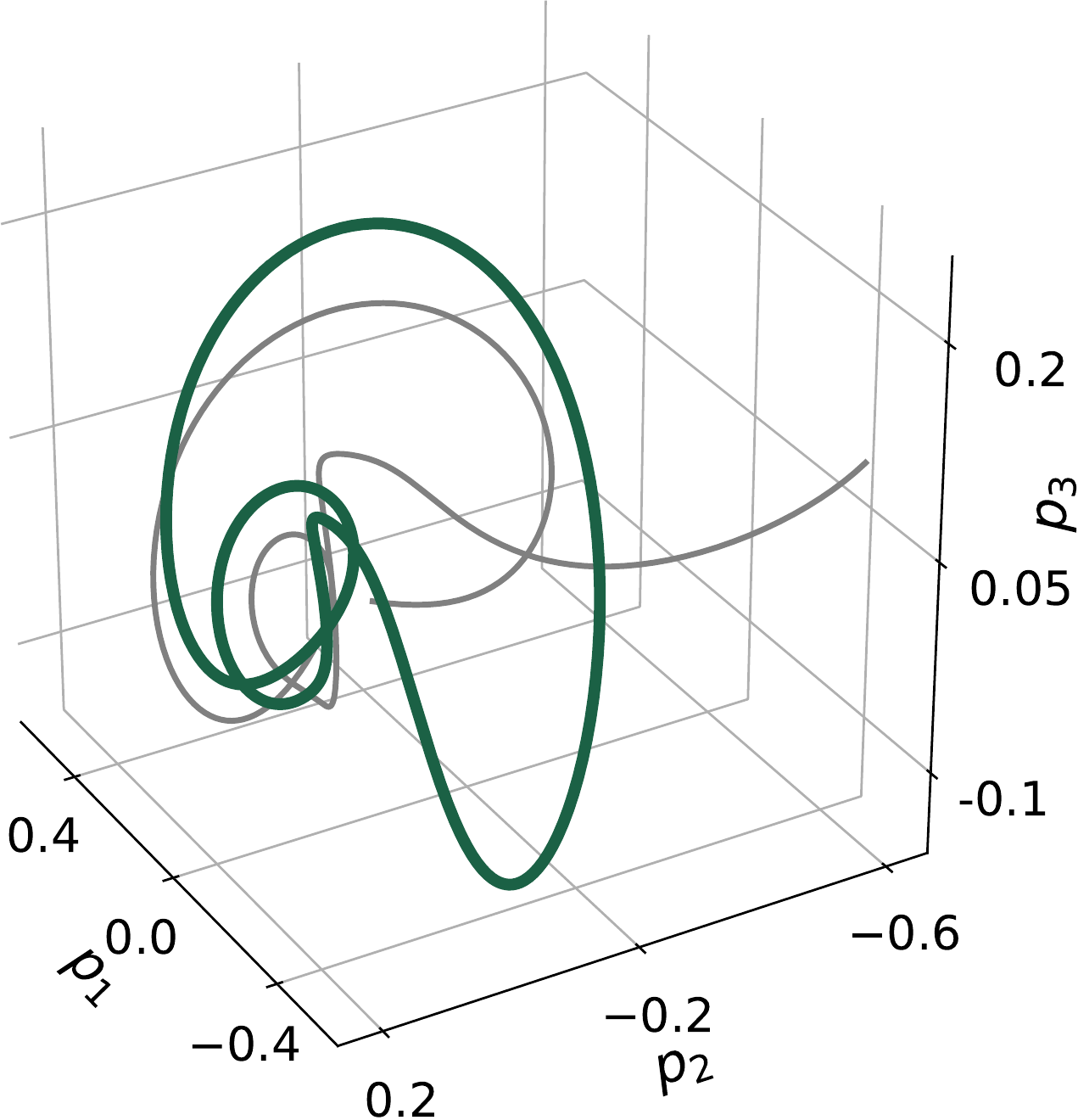}\,
	\subgraphicw{0.23}{e}{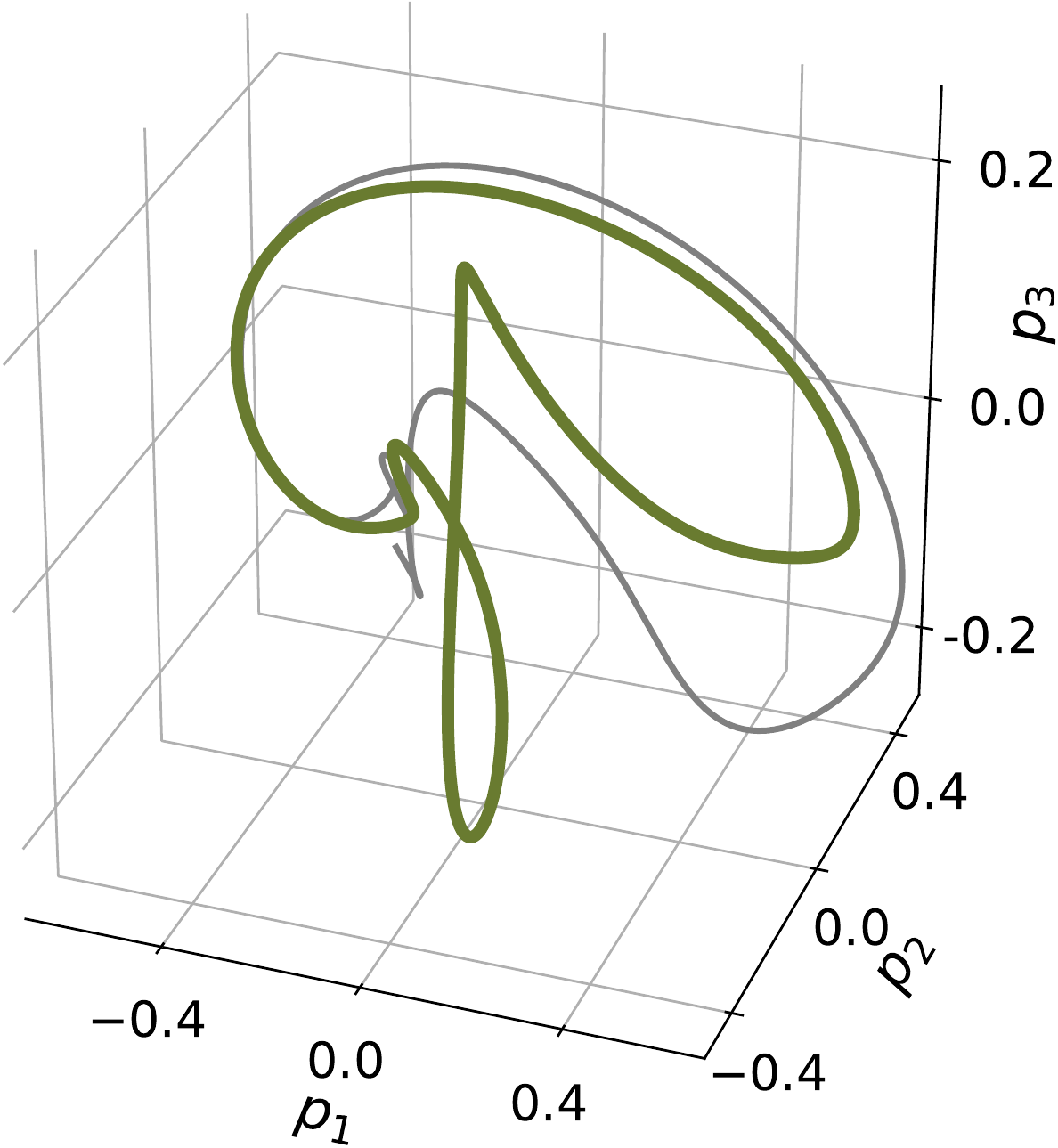}\,
	\subgraphicw{0.23}{f}{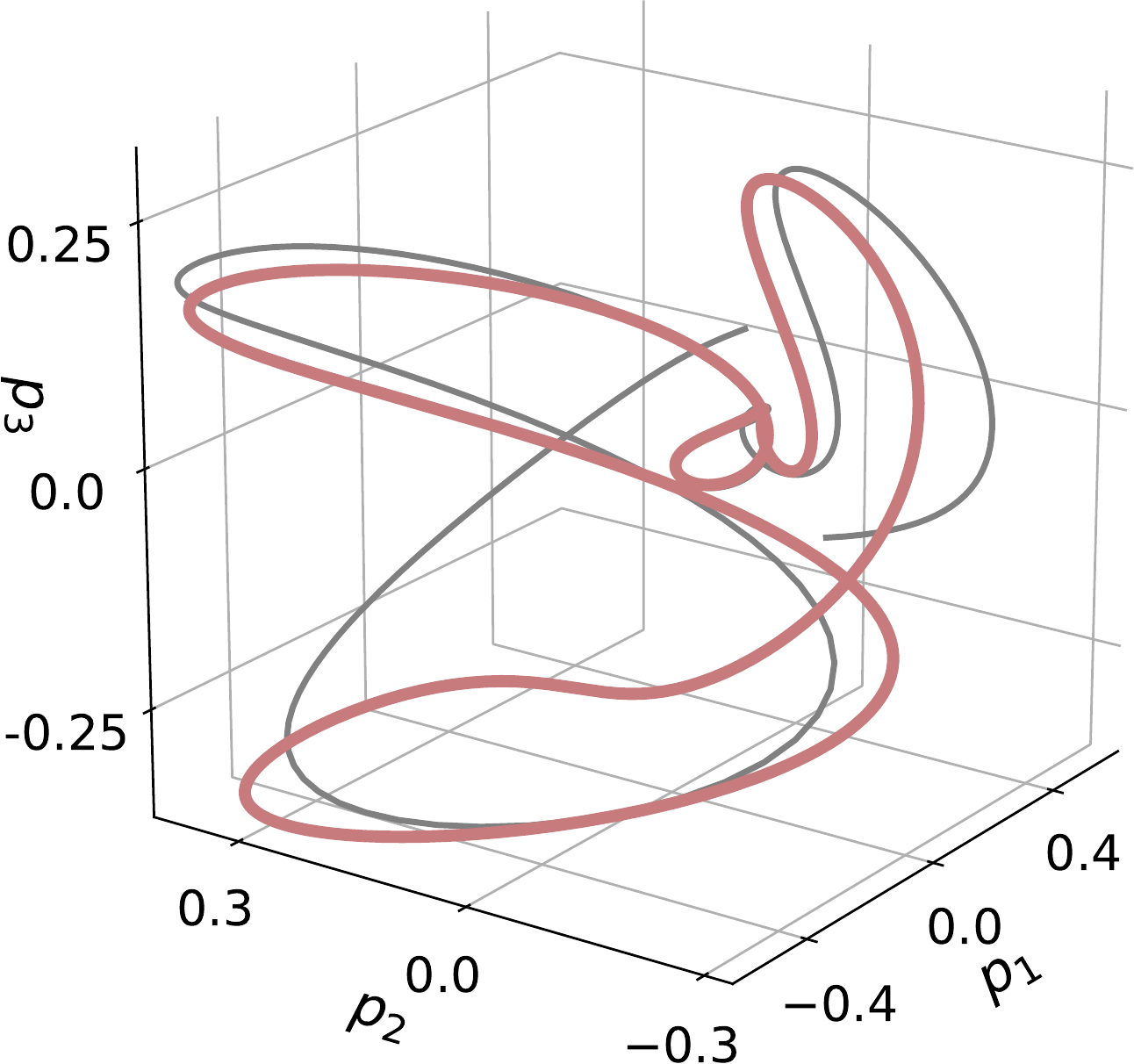} \\
	\subgraphich{0.24}{g}{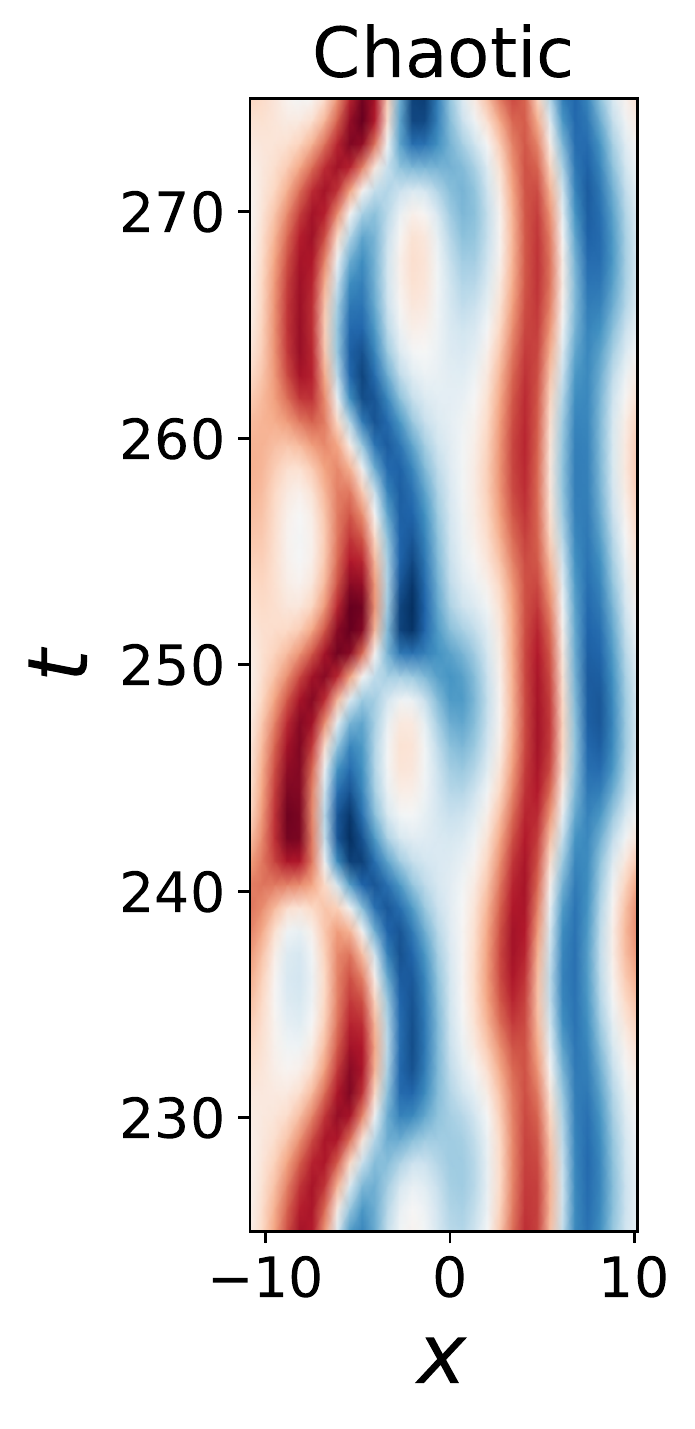}
	\includegraphics[height=0.24\textwidth]{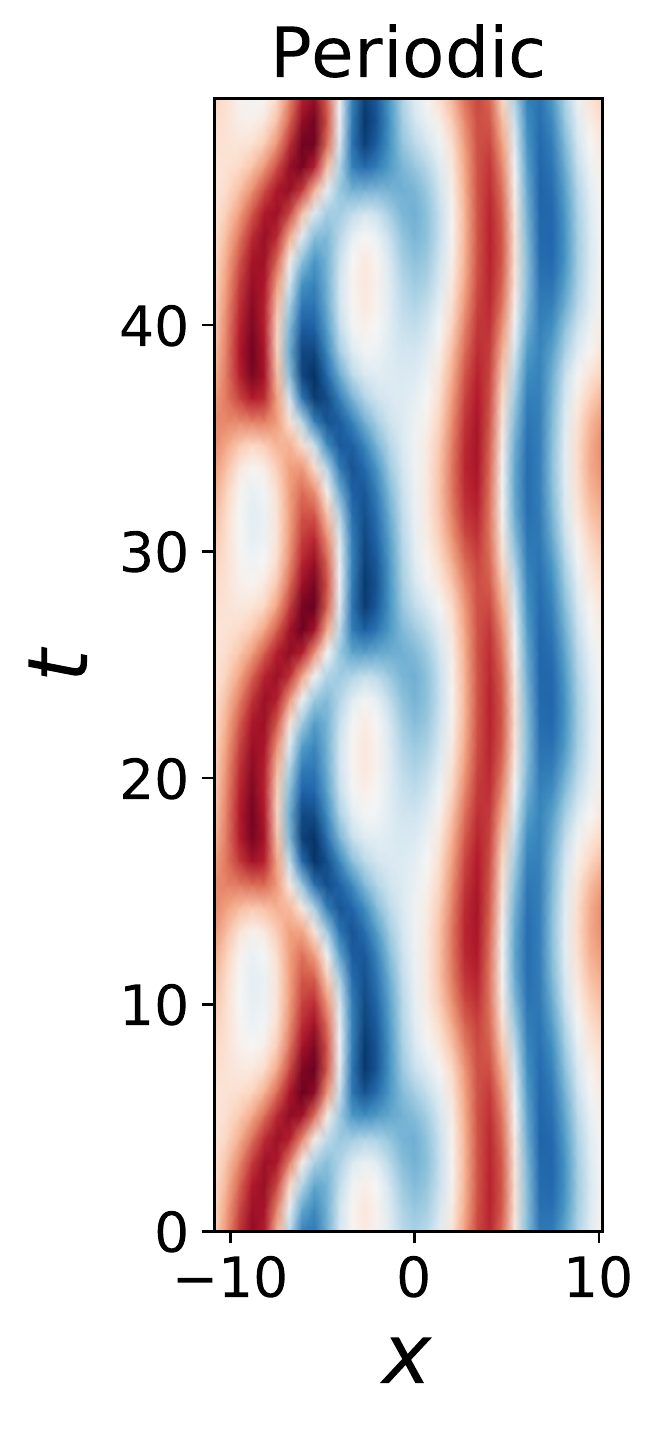}
	\subgraphich{0.24}{h}{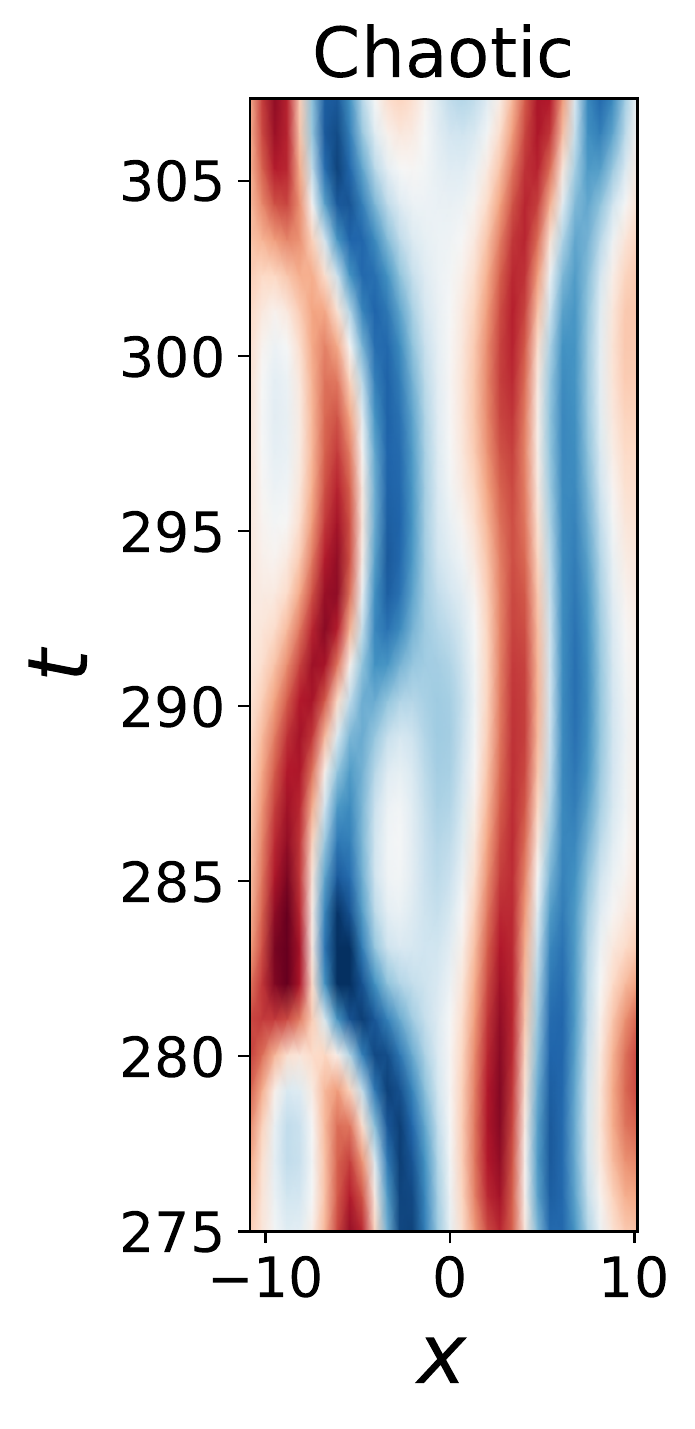}
	\includegraphics[height=0.24\textwidth]{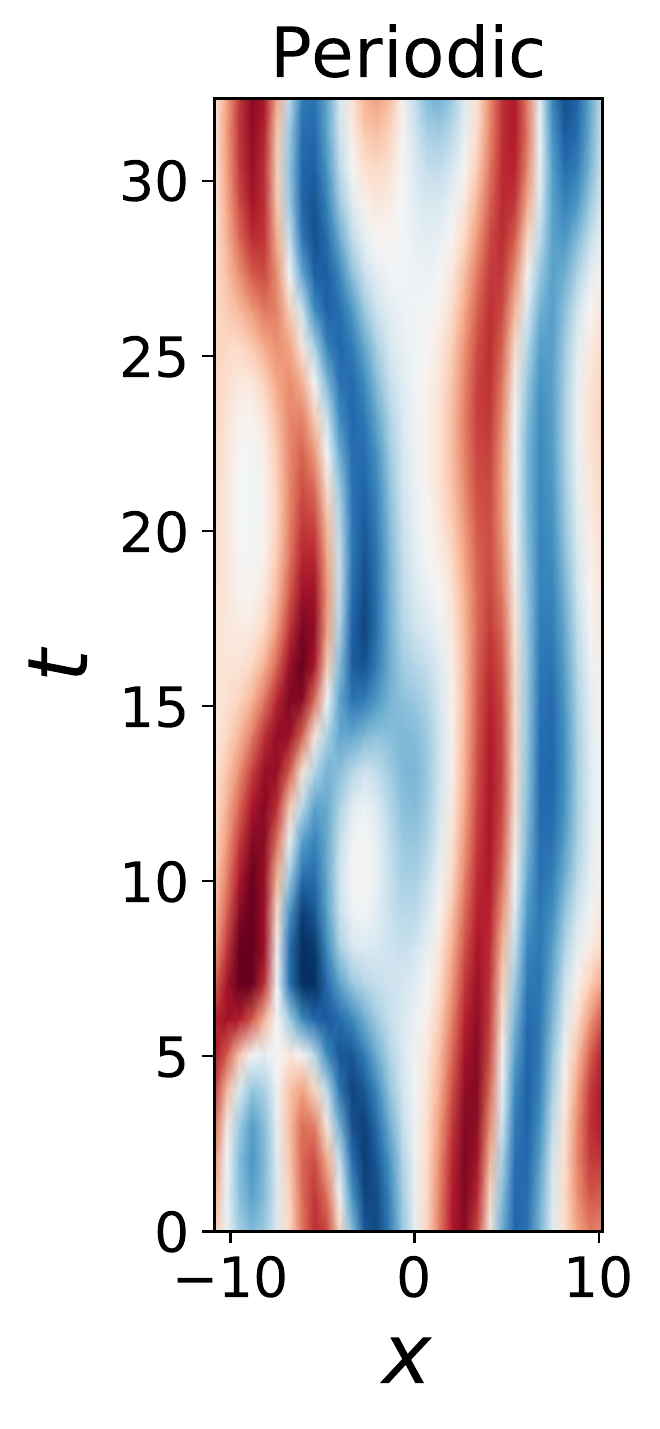}
	\subgraphich{0.24}{i}{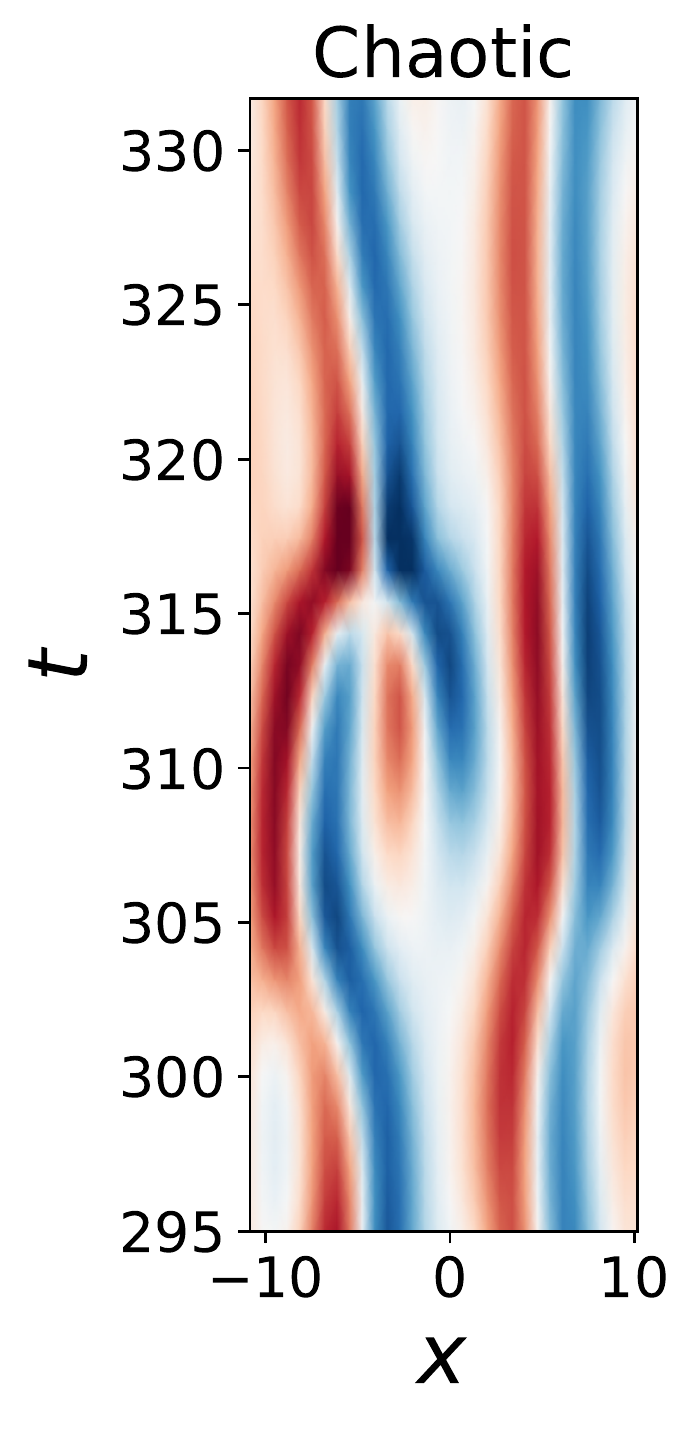}
	\includegraphics[height=0.24\textwidth]{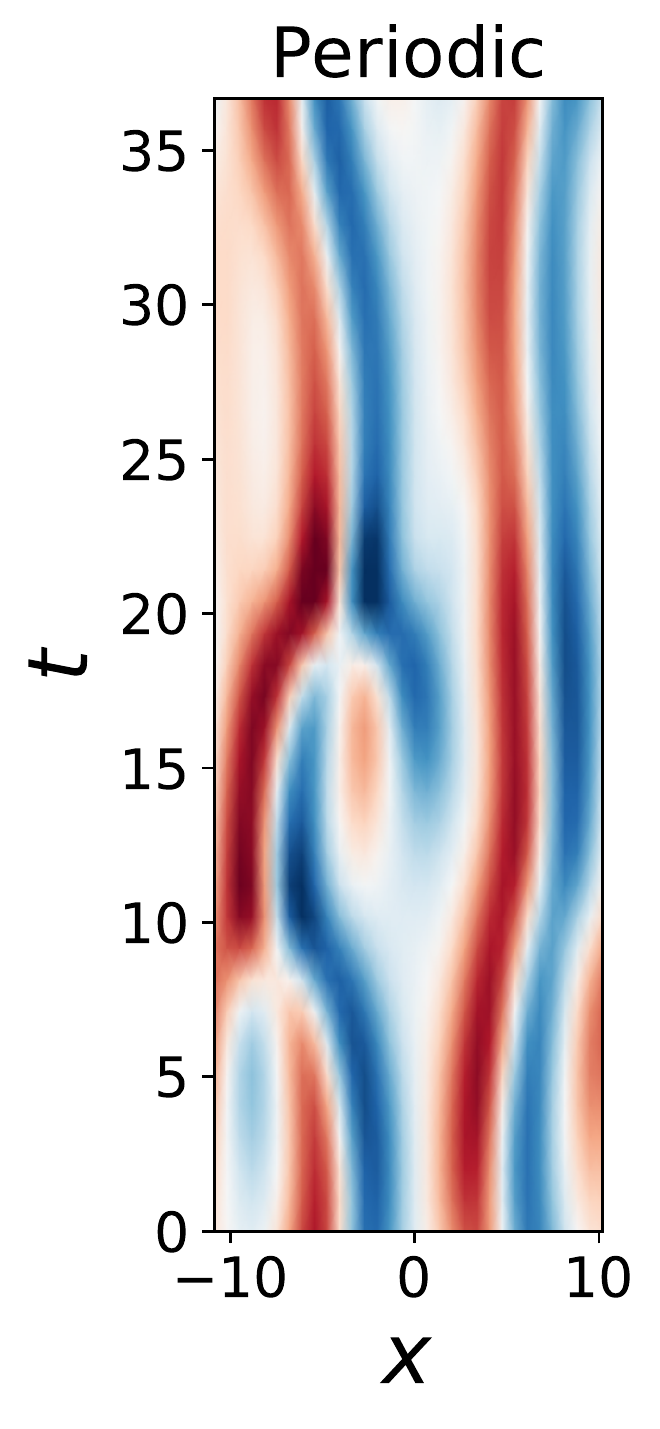}
	\subgraphich{0.24}{j}{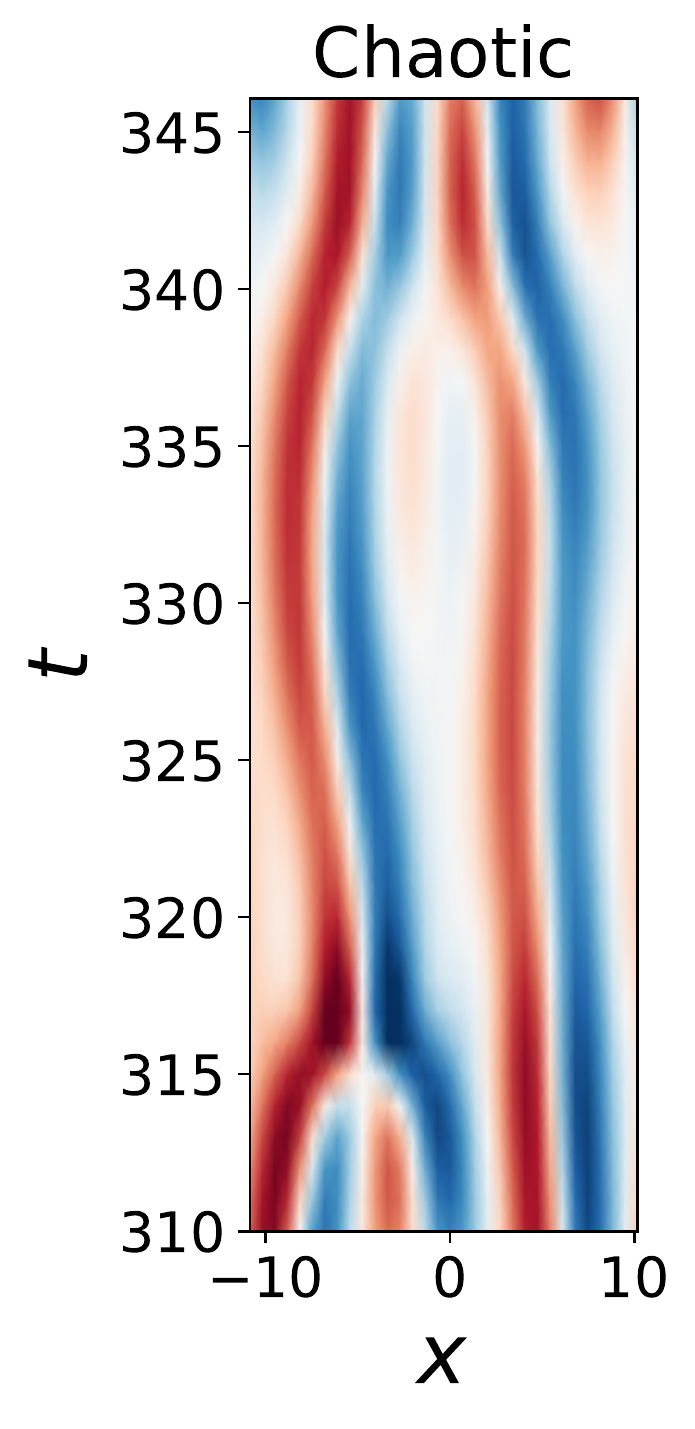}
	\includegraphics[height=0.24\textwidth]{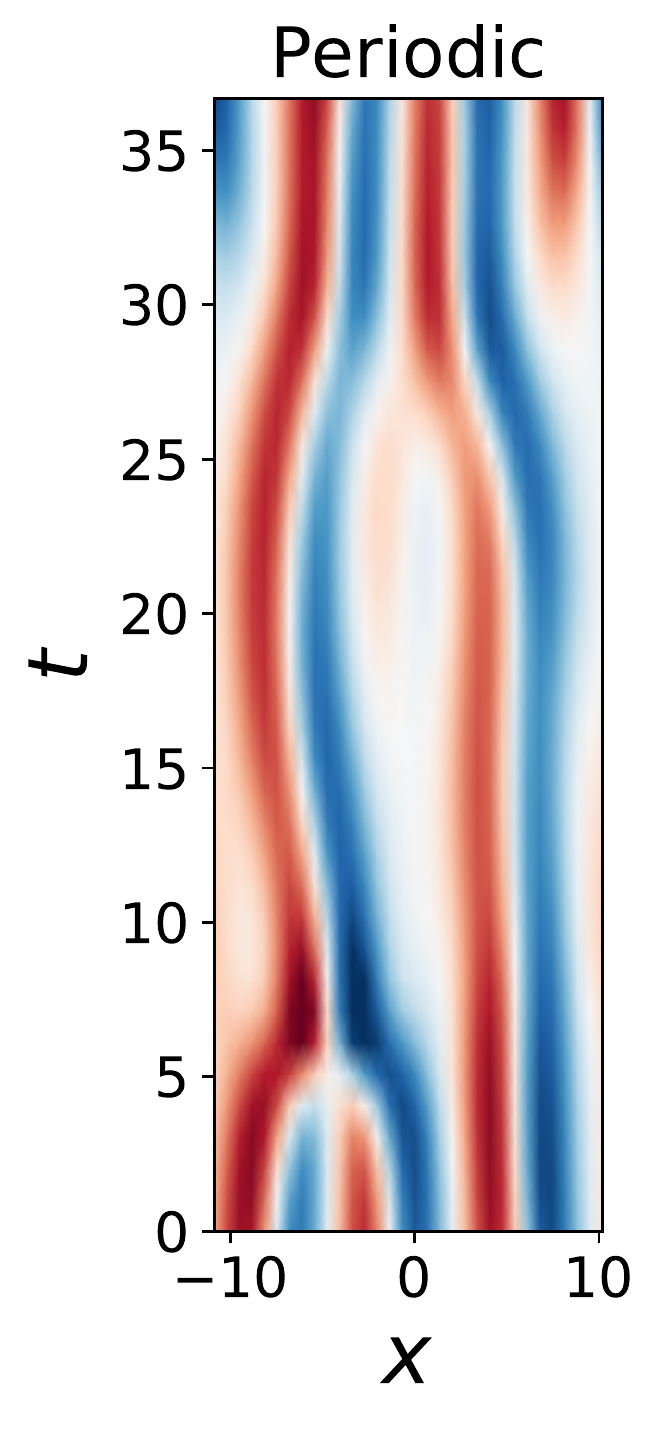} \\
	\includegraphics[width=0.25\textwidth]{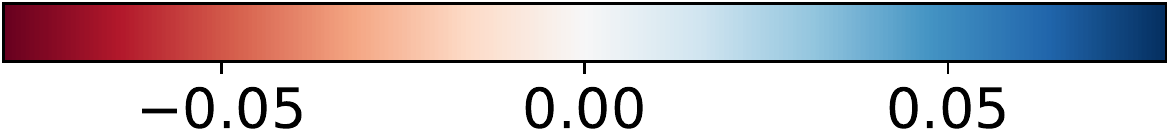}
	\caption{
		(a) Shadowing distances $\distShadow^{(i)}$
		of a chaotic trajectory to the periodic
		orbits $\po_i \,, i \in \{1, 2, 3, 4\}$
		of the \KS{} system. 
		(b) Zoom into the time interval of (a) that corresponds
				to the visualizations in (c--j).
		(c--f) Examples of shadowing trajectory segments which 
		correspond to the 
		time interval $\zeit \in [225, 345]$, shown in (b), 
		visualized as local projections 
		of chaotic trajectories segments (gray) onto the local PCA bases
		of the respective periodic orbits $\po_{1}, \ldots, \po_{4}$
		(colors/bold) 
		that are shadowed. 
		(g--j) Chaotic trajectory 
		segments next to the periodic solutions $\po_{1}, \ldots, \po_{4}$
		that they
		 shadow, visualized in space-time by color-coding 
		 	          the amplitude of $u(x, \zeit)$. 
		Time intervals of the shadowing 
		trajectory segments (g--j) in space-time visualizations are the same 
		intervals shown in the local projections (c--f).
 		\label{f-SSPAKS}}
\end{figure*}

We sample each orbit in our base set \refeq{e-baseset} with the 
constant sampling time $\zeit_s = 0.5$ and use these samples to 
generate local PCA bases and persistence diagrams for each 
periodic orbit as described in \refsect{e-persistPO}. 
In \reffig{f-SSPAKS} (a), we show the shadowing distances 
\refeq{e-distShadow} of a 
chaotic trajectory from the four periodic
orbits of the \KS{} system as a function of time.
Similar to \reffig{f-RosslerDistShadow}, we normalized 
each shadowing distance by its maximum.
In computing these 
distances, we used unit weights $w_0 = w_1 = 1$ in
\refeq{e-distShadow} and the Wasserstein distance \refeq{e-Wasserstein}
with $p = q = 2$. 

When the  \shadowDist{} \refeq{e-distShadow} to a particular
periodic orbit is small, we expect to find the chaotic trajectory 
segment to have a shape similar to that of the respective periodic
orbit. We illustrate this on the local 
projections of \reffig{f-SSPAKS} (c--f), where we show chaotic
trajectory segments of different durations with initial conditions
corresponding to the local minima of the shadowing distances in 
\reffig{f-SSPAKS} (b) along with the projections of the periodic
orbits. As further evidence, we show the 
space-time visualizations of the shadowing trajectory segments 
next to the periodic orbits in \reffig{f-SSPAKS} (g--j), where 
the amplitude of the scalar field $u(x, \zeit)$ is color-coded. 
The time interval shown in \reffig{f-SSPAKS} (g) spans approximately 
five periods of $\po_1$, whereas for the rest of the periodic
orbits one period is shown in 
\reffig{f-SSPAKS} (h--j). Note that since the space-time 
visualizations are not symmetry reduced, the relative periodic 
orbits arrive at a symmetry-transformed state after one period. 
This can easily be seen on \reffig{f-SSPAKS} (j), 
where the final 
state is the initial state, shifted in space by 
$\delta x \approx 12.07$. In \reffig{f-SSPAKS} (g--i), the initial
and final states are related by reflection.
 
\section{Discussion}
\label{s-discussion}
 
We presented the results of \perShadow{} in the R\"ossler 
and the \KS{}
systems where in both cases, we saw that the \shadowDist{} 
successfully 
captured the shape similarities between periodic orbits and 
chaotic trajectory segments. 
We chose the R\"ossler system for the first application of 
\perShadow{} in order to avoid technical difficulties
associated with PDEs and illustrate the core concepts of
our method on a simple example. 
With this in mind, we considered only two periodic orbits,
namely $\overline{1}$ and $\overline{01}$, in our analysis.
We would like to note the limitations that come with 
this choice: 
In general, the trajectories of the R\"ossler system 
can have itineraries containing the symbol 
sequence $00$ (see \reffig{f-RosslerMarkov} (a)), 
however, none of the periodic orbits $\overline{1}$ or
$\overline{01}$ contain this symbol sequence.
Thus, the base set of periodic orbits that we considered in 
\refsect{s-Roessler} cannot capture 
trajectory segments which are associated with the symbol sequence 
$00$.
We could have captured some these motions by 
including the periodic orbit $\overline{001}$ into our 
analysis, however, we refrained from doing so for
the clarity of the presentation.

The success
of \perShadow{} in the \KS{} system
reveals the true potential of the method for 
the study of high-dimensional systems.
It is important to note that in \reffig{f-SSPAKS} (a) at almost all 
times, at least one of the \shadowDist{}s is less than $0.5$
and has
a local minimum.
This demonstrates that the spatiotemporally chaotic
\KS{} dynamics can be approximated by a Markov chain
based on the four 
periodic solutions.
Notice also that some of the step-like minima
of the shadowing distances in \reffig{f-SSPAKS} (a) coincide:
For example, both $\distShadow^{(1)}$ and $\distShadow^{(2)}$
start at a low value with 
similar instances in the future. This suggests that
$\po_1$ and $\po_2$ could be related through a bifurcation.
Indeed, $\po_2$ appears
on the unstable manifold of $\po_1$ at a lower value of the control
parameter $L$ as demonstrated
in \refref{BudCvi15}. 
Another important observation to make on 
\reffig{f-SSPAKS} (a) is that all dips in $\distShadow^{(4)}$
are preceded by those of $\distShadow^{(3)}$. 
This suggests that 
$\po_3$ admits a symmetry-breaking instability since $\po_4$ has a
nonzero spatial drift, see \reffig{f-SSPAKS} (j). 
A detailed investigation and periodic-orbit-based modeling  
of the \KS{} dynamics will be a subject of a future study.

In order to test our method's robustness against the 
choice of 
norm, we partially repeated our calculations using 
``randomly modified''
norms. To this end, we defined
\begin{equation}
\inprod{\svec^{(i)}}{\svec^{(j)}}_{R^{(l)}} = 
\sum_{k=1}^{D} R^{(l)}_k \svec^{(i)}_{k} \svec^{(j)}_{k} \,, 
\label{e-inprod-random}
\end{equation}
where $R_k^{(l)}$ are positive pseudorandom numbers, the sum of which is 
equal to $D$, the system dimension. Note that if we choose 
$R_k = 1$, we recover the $L_2$-norm
\refeq{e-inprod}. We found that the shadowing distances
obtained with modified norms looked qualitatively similar to those 
in \reffig{f-RosslerDistShadow} and \reffig{f-SSPAKS}. 
Although it is beyond the scope of the present work, we 
speculate that the robustness of the \shadowDist{} 
\refeq{e-distShadow}
against the modifications \refeq{e-inprod-random} of 
the inner product 
could potentially be established rigorously, with techniques 
similar to those used to prove the stability of persistence 
diagrams.\rf{CEH2007}

In our analyses, we chose
the sampling time $t_s$, 
Wasserstein distance degree $p$,
and the \shadowDist{} weights $w_i$
through numerical experimentation.
We first produced the data for these experiments 
by
generating shadowing trajectories 
from the
initial conditions 
corresponding to slight perturbations to
periodic orbits as in \reffig{f-RosslerMarkov} (b). 
We then generated persistence diagrams and computed 
the associated \shadowDist{}s for different choices of 
parameters and then settled with the ones that 
yielded the expected shadowing signals.
We chose the sampling time $t_s$ via
a trade-off: 
When $\zeit_s$ was too 
long the persistence diagrams missed
significant topological
features of underlying trajectories; 
whereas when the sampling time was too short 
the persistence diagrams associated with the 
periodic orbits and shadowing trajectories 
differed significantly. 
In choosing the sampling time, 
we avoided both of these extremes.
Apart from the Wasserstein distance degree $p = 2$ that
we used, we also tried $p = 1$,
which resulted in shadowing signals with smoother variations in time.
Consequently, we decided to use $p = 2$, since sharper variations in 
time would be more amenable to shadowing detection with a 
thresholding algorithm. 
Besides the unit weights $w_0 = w_1 = 1$ in the
\shadowDist{} \refeq{e-distShadow}, we also tried 
$w_{0,1} = [W_p (\PerD_{0,1}^0 , \PerD_{0,1}^{\po_i})]^{-1}$, 
where $\PerD_{0,1}^0$ are trivial persistence diagrams with 
diagonal elements only. 
This choice overemphasized holes,
which resulted in the corresponding shadowing distance
time series missing some of the symbol assignments
in the R\"ossler system.
\section{Conclusion and perspectives}
\label{s-conclusion}

In this paper, we introduced \perShadow{} for 
inferring the 
symbolic dynamics of a chaotic time series by quantifying 
the shape similarity 
of chaotic trajectory segments and periodic solutions of 
the system. Our starting motivation was to have a tool for 
understanding high-dimensional chaos in terms of the periodic solutions 
of a system and we demonstrated that \perShadow{} can be utilized
for 
this purpose by successfully applying it to the spatiotemporally 
chaotic \KS{} system. We are now in position to apply our method to 
problems that are
computationally much more challenging, such as the 
simulations of the Navier--Stokes equations in three dimensions.

We would like to mention that our use of persistence is  
similar in spirit to the ``Sliding Windows and 1-dimensional
Persistence Scoring'' (\textit{SW1PerS}) method,\rf{PereaHarer2015}
in which one constructs delay embeddings of  
time series data before carrying the persistence computation in 
order to detect periodicities in the data. 
In our case, we do not need a delay embedding since we assume that
we have access to the complete state space information.
Moreover,
instead of trying to detect periodicities in the signal, 
we try 
to identify similarities to a certain precomputed set of periodic
solutions in \perShadow{}. One can imagine applications
in which the two methods are mixed. 
For example, if one is searching 
for shadowing in a laboratory experiment in which the complete 
state measurement is not available, \perShadow{} could be carried 
out on a delay embedding.
Another interesting 
hybrid application could be searching for periodic 
solutions using the state space persistence of time series data.

In this paper, we opted for the simplicity of 
the presentation rather 
than fine-tuning our tools. 
As a consequence, there are 
a lot of aspects of \perShadow{} that could potentially 
be optimized for different settings.
As we discussed in \refsect{s-discussion},
the parameters such as 
sampling time, 
Wasserstein distance degree, 
and shadowing distance weights
should be chosen according to the specific
properties of a problem.
Our exploration of this free parameter space was
by no means exhaustive and we expect that 
these
choices will need to be revisited when applying
\perShadow{} in different settings.
Establishing the guidelines for this purpose
will be a topic of our future research.

\begin{acknowledgments}
We are grateful to Predrag Cvitanovi\'c for his comments on 
an early version of this manuscript
and to the anonymous referee, 
whose suggestions helped us to 
improve this paper.
\end{acknowledgments}

\appendix
\section{State space of the Kuramoto--Sivashinsky system}
\label{a-symred}

We begin our numerical formulation by plugging the Fourier expansion
$u = \sum_k \Fu_k (\zeit) e^{i q_k x}$, where $q_k = 2 \pi k / L$ and
$k = \ldots, -2, -1, 0, 1, 2, \ldots$, into the \KSe{}
\refeq{e-ks} in order to obtain the infinite set of ODEs
\beq
\dot{\Fu}_k = ( q_k^2 - q_k^4 )\, \Fu_k
- i \frac{q_k}{2} \!\sum_{m=-\infty}^{+\infty} \!\!\Fu_m \Fu_{k-m}
\,.
\label{e-Fks}
\eeq
Noting that the $0^\text{th}$
Fourier mode $\Fu_0$ is decoupled from 
the rest and $\Fu_{-k} = \Fu_k^*$ due to the realness of 
$u (x, \zeit)$, a truncated state vector of the \KS{} system 
can be expressed as 
\beq
	\svec = (a_1, b_1, a_2, b_2, \ldots, a_N, b_N) \,, 
	\label{e-svecKS}
\eeq
where $(a_k, b_k) = (\Re \Fu_k, \Im \Fu_k)$ and $N$ is the highest
Fourier mode that is kept in the expansion. In our computations,
we used $N=15$, the adequacy of which was demonstrated in \refref{SCD07}. 
In our codes, the nonlinear term in \refeq{e-Fks} is computed 
pseudospectrally\rf{Canuto2007} and the time-stepping is carried 
out using the general-purpose integrator \texttt{odeint} from
\texttt{scipy},\rf{scipy} which itself is a wrapper of
\texttt{lsoda} from the \texttt{ODEPACK} library.\rf{hindmarsh1983}

It is straightforward to confirm that 
the action of the symmetries \refeq{e-gx} and 
\refeq{e-sigma} on the real-valued state space coordinates 
\refeq{e-svecKS} 
are
\beq
	\LieEl_x (\delta x) (a_k, b_k) = R(- k \phi) (a_k, b_k) 
\eeq
and
\beq
	\sigma (a_k, b_k) = (-a_k, b_k)\,, \label{e-sigmaF}
\eeq
where $\phi = 2 \pi \, \delta x / L$ and $R (\theta)$ is the 
$2 \times 2$ rotation matrix
\beq
	R(\theta) =  \begin{pmatrix}
		\cos \theta & - \sin \theta   \\
		\sin \theta &   \cos \theta
	\end{pmatrix} \, .
\eeq
The \fFslice{} method of \refref{BudCvi14} fixes the polar angle on 
the subspace spanned by the first Fourier mode, \ie\ $(a_1, b_1)$, 
in order to eliminate the spatial drifts. It was already 
demonstrated in \refref{BudCvi14} that such a transformation in the
\KS{} system  leads to rapid fluctuations in time, which, in general,
could be regularized by rescaling the time variable. Here, we tackle 
this problem by a different approach, which we found simpler to use 
in \perShadow{}. Let $\svec$ be a generic state of the \KS{} system
with nonzero components in the second Fourier mode subspace, \ie\
$a_2^2 + b_2^2 > 0$. We search for a shifted state 
\beq
	\gamma = \LieEl_x (- \delta \hat{x}) \svec \label{e-svec2ndF}
\eeq
such that  
\(
	\gamma = (\hat{a}_1', \hat{b}_1', \hat{a}_2', \hat{b}_2', 
				 \ldots )
\)
has
\beq
\hat{a}_2' = 0,\,
\hat{b}_2' > 0 \,. \label{e-2ndFcond}
\eeq
Transforming to $\gamma$ \refeq{e-svec2ndF} eliminates the 
continuous translation degree of freedom by fixing the phase of the 
second Fourier mode. However, it does not fully reduce this symmetry 
since if $\gamma$ \refeq{e-svec2ndF} satisfies \refeq{e-2ndFcond}
so does 
\beq
\LieEl_x (L/2) \gamma =  (- \hat{a}_1', - \hat{b}_1', 
0, \hat{b}_2', 
- \hat{a}_3', - \hat{b}_3',
\hat{a}_4', \hat{b}_4',
\ldots )\,. \label{e-gxhalfL}
\eeq 
In other words, transformation 
to \refeq{e-svec2ndF} turns the continuous translation symmetry into 
a discrete one. As we shall see, this discrete symmetry can be 
reduced by the construction of invariant polynomials similar to those 
introduced in \refref{BudCvi15}.

After the transformation \refeq{e-svec2ndF}, the state space has 
two discrete symmetries, whose actions flip the signs of a subset of 
the state space coordinates. Notice that the action of the reflection
$\sigma$ \refeq{e-sigmaF} does not break the condition 
\refeq{e-2ndFcond} since $\svecRed'$ has $a_2' = 0$ and $b_2$ is 
invariant under $\sigma$. Following the recipe of \refref{BudCvi15},
we can define a reflection-reduced state vector as 
\bea
	\rho &=& (a_1'^2 - a_3'^2, 
				    b_1', b_2', 
				    a_1' a_3', 
				    b_3', 
				    a_3' a_4', 
				    b_4', 
				    a_4' a_5',
				    b_5' \continue 
				&&  a_5' a_6',
				    b_6', 
				    a_6' a_7',
				    b_7',  
				    a_7' a_8',
				    b_8' \ldots)\,, \label{e-svecRefRed}
\eea
where we omitted $a_2'$, since it is set to $0$. Note that 
\refeq{e-svecRefRed} is invariant under the sign change of all 
$a_k$ and not invariant under the sign change of any other 
subset of $a_k$s.

We can now turn our attention to the
discrete symmetry due to the half-domain shift \refeq{e-gxhalfL}. 
We should first find the representation of this symmetry on the 
reflection-invariant polynomial coordinates \refeq{e-svecRefRed}. 
Denoting the $k^\text{th}$
elements of \refeq{e-svecRefRed} by 
$\rho_k$ it follows from inspection that
\bea
	\LieEl_x (L/2) \rho &=& 
						    (\rho_1, 
  						   - \rho_2, 
						     \rho_3,
					    	 \rho_4,
					       - \rho_5,
				    	   - \rho_6,
							 \rho_7,
						   - \rho_8,
						   - \rho_9,\continue 
						&& - \rho_{10},
							 \rho_{11},
						   - \rho_{12},
						   - \rho_{13},
						   - \rho_{14},
							 \rho_{15} \ldots ) \, .
								 \label{e-LieElsvecreddd}
\eea
Beginning with $\rho_8$, every element of $\rho$ except 
$\{\rho_{11}, \rho_{15}, \rho_{19}, \rho_{23}, \rho_{27}, \ldots \}$
(every fourth element) changes its sign under the action of 
$\LieEl_x (L/2)$. Thus, we can write the final invariant polynomial
coordinates as 
\bea
	\svecRed &=& (\rho_1, 
				  \rho_2^2 - \rho_5^2, 
				  \rho_3, 
				  \rho_4,
				  \rho_2 \rho_5, 
				  \rho_5 \rho_6,
				  \rho_7,
				  \rho_6 \rho_8,  
			 	  \rho_8 \rho_9, 
			      \rho_9 \rho_{10}, \continue
			&&	  \rho_{11},
				  \rho_{10} \rho_{12}, 
			      \rho_{12} \rho_{13},
				  \rho_{13} \rho_{14}, 
				  \rho_{15},\, \ldots) \,. \label{e-svecRed}
\eea
While it might appear complicated,
the invariant polynomial 
coordinates \refeq{e-svecRefRed} and \refeq{e-svecRed} follow 
a regular pattern, and thus, are straightforward to implement. 
We used the symmetry-invariant state space coordinates 
\refeq{e-svecRed} to obtain the results of \refsect{s-KS}.

\bibliography{../../bibtex/neubauten.bib}{}
\bibliographystyle{unsrt}
\end{document}